\documentclass[a4paper,11pt]{article}
\usepackage{amssymb}
\usepackage{amsfonts}
\usepackage{graphicx}
\usepackage{epstopdf}
\usepackage{dcolumn}
\usepackage{amsmath}
\usepackage{hyperref}
\usepackage{latexsym,bm}
\usepackage{geometry}

\def \be {\begin{equation}}
\def \ee {\end{equation}}
\def \bea {\begin{eqnarray}}
\def \eea {\end{eqnarray}}
\def \nn {\nonumber}

\def \a {\alpha}
\def \b {\beta}
\def \g {\gamma}

\def \d {\delta}

\def \m {\mu}
\def \n {\nu}
\def \k {\kappa}

\def \s {\sigma}
\def \r {\rho}
\def \o {\omega}

\def \th {\theta}
\def \Th {\Theta}

\def \t {\tau}
\def \dag {\dagger}
\def \p {\partial}

\def\bd{\begin{document}}
\def\ed{\end{document}}
\def\nn{\nonumber}
\def\bea{\begin{eqnarray}}
\def\eea{\end{eqnarray}}
\let\bm=\bibitem
\let\la=\label

\def\N{{\cal N}}
\def\sst{\scriptscriptstyle}
\def\thetabar{\bar\theta}
\def\Tr{{\rm Tr}}
\def\one{\mbox{1 \kern-.59em {\rm l}}}

\def\a{\alpha}      \def\da{{\dot\alpha}}
\def\b{\beta}       \def\db{{\dot\beta}}
\def\c{\gamma}  \def\C{\Gamma}  \def\cdt{\dot\gamma}
\def\d{\delta}  \def\D{\Delta}  \def\ddt{\dot\delta}
\def\e{\epsilon}        \def\vare{\varepsilon}
\def\f{\phi}    \def\F{\Phi}    \def\vvf{\f}
\def\h{\eta}
\def\k{\kappa}
\def\l{\lambda} \def\L{\Lambda}
\def\m{\mu} \def\n{\nu}
\def\o{\omega}
\def\P{\Pi}
\def\r{\rho}
\def\s{\sigma}  \def\S{\Sigma}
\def\t{\tau}
\def\th{\theta} \def\Th{\Theta} \def\vth{\vartheta}
\def\X{\Xeta}
\def\z{\zeta}
\def\w{\wedge}
\def\u{\underline}
\def\hs{\hspace}

\def\cA{{\cal A}} \def\cB{{\cal B}} \def\cC{{\cal C}}
\def\cD{{\cal D}} \def\cE{{\cal E}} \def\cF{{\cal F}}
\def\cG{{\cal G}} \def\cH{{\cal H}} \def\cI{{\cal I}}
\def\cJ{{\cal J}} \def\cK{{\cal K}} \def\cL{{\cal L}}
\def\cM{{\cal M}} \def\cN{{\cal N}} \def\cO{{\cal O}}
\def\cP{{\cal P}} \def\cQ{{\cal Q}} \def\cR{{\cal R}}
\def\cS{{\cal S}} \def\cT{{\cal T}} \def\cU{{\cal U}}
\def\cV{{\cal V}} \def\cW{{\cal W}} \def\cX{{\cal X}}
\def\cY{{\cal Y}} \def\cZ{{\cal Z}}

\def\ua{\underline{\alpha}} \def\ubb{\underline{\beta}}
\def\ug{\underline{\gamma}}
\def\ub{\underline{\phantom{\alpha}}\!\!\!\beta}
\def\uc{\underline{\phantom{\alpha}}\!\!\!\gamma}
\def\um{\underline{\mu}} \def\un{\underline{\nu}}
\def\ud{\underline\delta}
\def\ue{\underline\epsilon}
\def\una{\underline a}\def\unA{\underline A}
\def\unb{\underline b}\def\unB{\underline B}
\def\unc{\underline c}\def\unC{\underline C}
\def\und{\underline d}\def\unD{\underline D}
\def\une{\underline e}\def\unE{\underline E}
\def\unf{\underline{\phantom{e}}\!\!\!\! f}\def\unF{\underline F}
\def\unm{\underline m}\def\unM{\underline M}
\def\unn{\underline n}\def\unN{\underline N}
\def\unp{\underline{\phantom{a}}\!\!\! p}\def\unP{\underline P}
\def\unq{\underline{\phantom{a}}\!\!\! q}
\def\unQ{\underline{\phantom{A}}\!\!\!\! Q}
\def\unH{\underline{H}}
\def\ul{\underline}

\def\As {{A \hspace{-6.4pt} \slash}\;}
\def\bs {{b \hspace{-6.4pt} \slash}\;}
\def\Ds {{D \hspace{-6.4pt} \slash}\;}
\def\ds {{\del \hspace{-6.4pt} \slash}\;}
\def\ss {{\s \hspace{-6.4pt} \slash}\;}
\def\ks {{ k \hspace{-6.4pt} \slash}\;}
\def\ps {{p \hspace{-6.4pt} \slash}\;}
\def\pas {{{p_1} \hspace{-6.4pt} \slash}\;}
\def\pbs {{{p_2} \hspace{-6.4pt} \slash}\;}

\def\Fh{\hat{F}}
\def\Vh{\hat{V}}
\def\Xh{\hat{X}}
\def\ah{\hat{a}}
\def\xh{\hat{x}}
\def\yh{\hat{y}}
\def\ph{\hat{p}}
\def\xih{\hat{\xi}}

\def\psit{\tilde{\psi}}
\def\Psit{\tilde{\Psi}}
\def\tht{\tilde{\th}}

\def\At{\tilde{A}}
\def\Qt{\tilde{Q}}
\def\Rt{\tilde{R}}
\def\Nt{\tilde{N}}

\def\at{\tilde{a}}
\def\st{\tilde{s}}
\def\ft{\tilde{f}}
\def\pt{\tilde{p}}
\def\qt{\tilde{q}}
\def\vt{\tilde{v}}
\def\nt{\tilde{n}}

\def\delb{\bar{\partial}}
\def\bz{\bar{z}}
\def\bD{\bar{D}}
\def\bB{\bar{B}}

\def\bk{{\bf k}}
\def\bl{{\bf l}}
\def\bp{{\bf p}}
\def\bq{{\bf q}}
\def\br{{\bf r}}
\def\bx{{\bf x}}
\def\by{{\bf y}}
\def\bR{{\bf R}}
\def\bV{{\bf V}}

\def\d{\delta}\def\D{\Delta}\def\ddt{\dot\delta}

\def\p{\partial} \def\del{\partial}
\def\xx{\times}
\def\uno{\mbox{1 \kern-.59em {\rm l}}}

\def\trp{^{\top}}
\def\inv{^{-1}}
\def\dag{{^{\dagger}}}

\def\pr{\prime}
\def\rar{\rightarrow}
\def\lar{\leftarrow}
\def\lrar{\leftrightarrow}

\title{Higher Spin Entanglement Entropy}
\author{
Jiang Long\footnote{lj301@pku.edu.cn}
}
\date{}
\begin{document}
\maketitle
\begin{center}
{{\it
Department of Physics and State Key Laboratory of Nuclear Physics and Technology, Peking University, No. 5 Yiheyuan Rd, Beijing 100871, P.R.\! China
\vspace{2mm}
}}
\vspace{10mm}
\end{center}

\date{}
\begin{abstract}
In this paper, we develop a perturbation formulation to calculate the single interval higher spin R$\acute{e}$nyi and entanglement entropy for two dimensional conformal field theory with $\mathcal{W}_{\infty}(\lambda)$ symmetry. The system is at finite temperature and is deformed by higher spin chemical potential. We manage to compute higher spin R$\acute{e}$nyi entropy with various spin deformations up to order $\mathcal{O}(\mu^2)$. For spin 3 deformation, we calculate exact higher spin R$\acute{e}$nyi and entanglement entropy up to $\mathcal{O}(\mu^4)$.  When $\lambda=3$, in the large $c$ limit, we find perfect match with tree level holographic higher spin entanglement entropy up to order $\mu^4$ obtained by the Wilson line prescription. We also find quantum corrections to higher spin entanglement entropy which is beyond tree level holographic results. The quantum correction is universal at order $\mu^4$ in the sense that it is independent of $\lambda$. Our computation relies on a multi-valued conformal map from $n$-sheeted Riemann surface $\mathcal{R}_n$ to complex plane and correlation functions of primary fields on complex plane.  The method can be applied to general conformal field theories with $\mathcal{W}$ symmetry.
 \end{abstract}
 \newpage
\section{Introduction}
Entanglement entropy and its generalization R$\acute{e}$nyi entropy are important quantities to study quantum systems. Entanglement entropy is a good parameter to characterize the effective degree of freedom of a region which is entangled with the rest of the system. It has interesting applications in condensed matter systems \cite{Calabres05,Kitaev0501,Levin0510}. In general, they are hard to compute. Surprisingly, inspired by $AdS/CFT$ correspondence, Ryu and Takayanagi \cite{Ryu05} proposed that entanglement entropy in a conformal field theory could be calculated from a minimal surface in the dual bulk. Their beautiful work provides an effective way to compute entanglement entropy. At the same time, it opens a new window to study $AdS/CFT$, especially the emergence of spacetime in gravity. 

On the other side, another important development in $AdS/CFT$ in recent years is the higher spin holography \cite{sundborg01,Mikhailov,Klevanov02}, which suggests explicit duality between Vasiliev higher spin theory \cite{Vasi90,Vasi92} and vectorial conformal field theory in the large $N$ limit. Among various proposals, $HS/CFT_2$ \cite{Gaberdial}, which relates higher spin theory in $AdS_3$ \cite{3dVasiliev} to $\mathcal{W}_N$ minimal model in large $N$ limit, is of particular interest. This duality triggered various studies in $AdS_3$ gravity and CFT with $\mathcal{W}$ symmetry, please find nice reviews \cite{Gaberdiel13, Ammon1208} on this topic.

Higher spin R$\acute{e}$nyi entropy (HSRE) and higher spin entanglement entropy(HSEE) arise from the combination of the previous two separate branches. In the field theory side, when there are higher spin deformations, the partition function can be written schematically as
 \be
 Z=<\exp{-\mu\int W-\bar{\mu}\int\bar{W}}>,
  \ee
  where $\mu(\bar{\mu})$ is the chemical potential, $W(\bar{W})$is the corresponding higher spin current. It would be interesting to consider the deformation of entanglement entropy in this case. In the bulk, as the holographic entanglement entropy relates to a geometric object in usual Einstein-Hilbert theory, a generalization to holographic HSEE may provide insights on mysterious higher spin geometry. Also, they are expected to provide non-trivial check of $HS/CFT$ correspondence. Several works on this issue have been done. In \cite{bin1312,Perlmutter:2013paa}, for the theory with $\mathcal{W}$ symmetry, two interval R$\acute{e}$nyi entropy without classical higher spin deformation has been calculated in the short interval limit, the quantum one-loop results in the gravity and CFT side match exactly up to $\mathcal{O}(x^8)$, where $x$ is the cross ratio constructed from the two intervals. When there are higher spin chemical potential deformations, one interesting configuration\footnote{There are other configurations, such as chiral deformation solution \cite{chiraldef, probindbh}.} is the higher spin black hole \cite{kraus1103}. In these cases, a holographic HSEE has been proposed by Wilson line prescription \cite{Ammon1310, Boer1306}. It has been checked up to $\mathcal{O}(\mu^2)$ for the CFT with $\mathcal{W}_3$ symmetry \cite{Datta1406,Datta1405}. Other works related to holographic HSEE can be found in \cite{Datta}.

In this paper, we develop a general prescription to calculate single interval HSRE and HSEE from CFT side perturbatively. Instead of inserting twist operators, we use a multi-valued conformal map from a $n$-sheeted Riemann surface to a complex plane. For spin 3 deformation, we find the HSRE and HSEE of $\mathcal{W}_{\infty}(\lambda)$ theory up to $\mathcal{O}(\mu^4)$. We show that the $\mathcal{O}(\mu^2)$ correction of HSRE and HSEE are indeed universal, in the sense that they are independent of the value of $\lambda$ up to a normalization constant. At the $\mathcal{O}(\mu^4)$, we find the classical and all loop quantum corrections to HSRE and HSEE. For $\lambda=3$, the classical part matches with the gravity result exactly. Our method is sufficiently general to extend to all kinds of higher spin deformations.

The structure of this paper is as follows. In section 2, we review the general method to calculate N interval R$\acute{e}$nyi entropy and entanglement entropy in two dimensional conformal field theory, emphasizing the importance of a conformal map from $n$-sheeted Riemann surface to complex plane and its multi-valued property. In section 3, we review pure higher spin gravity in AdS$_3$, including its asymptotic symmetry, higher spin black hole and its partition function. Some important results on $\mathcal{W}$ symmetry and black hole partition function are shown at the same time. In section 4, we briefly review the holographic HSEE and expand the results of $sl(3)$ theory up to $\mathcal{O}(\mu^4)$. In section 5, we calculate perturbative partition function of higher spin black hole, which is a good exercise for HSRE and HSEE. In section 6, we calculate HSRE and HSEE up to $\mathcal{O}(\mu^4)$ and match HSEE to holographic results in section 4.  Discussion and conclusion are included in the last section. Some technical details are collected in four appendices.

\section{R$\acute{e}$nyi and Entanglement Entropy}
To define entanglement entropy, we suppose the system has a density matrix $\rho$ and then divide the system into a subsystem A and its complement B. The total Hilbert space is factorized into $\mathcal{H}_A\otimes\mathcal{H}_B$. We trace out the information of B and obtain the reduced density matrix of A,
\be
\rho_{A}=tr_B\rho,
\ee
then the entanglement entropy of A is defined as the standard von Neumann entropy,
\be
S_A=-tr_A\rho_A\log\rho_A.
\ee
A useful notion associated to the entanglement entropy is the $n$-th R$\acute{e}$nyi entropy $S^{(n)}$,
\be
S_A^{(n)}=\frac{1}{1-n}\log tr \rho_A^n=\frac{\log Z_n-n\log Z_1}{1-n}.\label{hsdf}
\ee
By analytical continuation of $n\to 1$, we find the entanglement entropy,
\be
S_A=\lim_{n\to1}S_A^{(n)}.
\ee
The procedure by introducing R$\acute{e}$nyi entropy to find the entanglement entropy is  called the replica trick \cite{Holzhey94}.

In two dimensions, the replica trick  is applied as follows. Suppose the original CFT is defined on a Riemann surface\footnote{Usually, this is the complex plane $\mathcal{C}$ or those can be conformally transformed to the complex plane. Indeed, in this paper, we will consider the complex plane or the finite temperature version, say, a cylinder.} $\mathcal{R}_1$. We uplift the theory to $n$ disconnected sheets and there is a branch cut along $A$ in each sheet. We glue the branch cut successively and find an $n$-sheeted Riemann surface  $\mathcal{R}_n$. The partition function in the $n$-sheeted Riemann surface is denoted as $Z_n$.

In general, the subsystem $A$ consists of $N$ intervals,
\be
A=\{\omega|Im\omega=0,Re\omega\in[u_1,v_1]\cup\cdots\cup[u_N,v_N]\}.
\ee
The partition function\footnote{We add a subscript $N$ to denote the number of intervals. Correspondingly, the $n$-sheeted Riemann surface is now denoted as $\mathcal{R}_{n,N}$.} $Z_{n,N}$ could be determined by introducing primary twist(and anti-twist) operators \cite{Calabres0406} with conformal dimension $h_n=\bar{h}_n=\frac{c}{24}(n-\frac{1}{n})$ at each point $u_i(v_i)$, such that $Z_{n,N}$ is a $2n$ point function of a $\mathbb{Z}_n$ cyclic orbifold conformal field theory on $\mathcal{R}_1$,
\be
Z_{n,N}=<1>_{\mathcal{R}_{n,N}}=<\sigma_n(u_1)\tilde{\sigma}_n(v_1)\cdots\sigma_n(u_N)\tilde{\sigma}_n(v_N)>_{\mathcal{R}_1}.\label{zn}
\ee
The expectation value of arbitrary operator $\mathcal{O}$ on $\mathcal{R}_{n,N}$ is
\be
<\mathcal{O}(z)>_{\mathcal{R}_{n,N}}=\frac{<\mathcal{O}(z)\sigma_n(u_1)\tilde{\sigma}_n(v_1)\cdots\sigma_n(u_N)\tilde{\sigma}_n(v_N)>_{\mathcal{R}_1}}{<\sigma_n(u_1)\tilde{\sigma}_n(v_1)\cdots\sigma_n(u_N)\tilde{\sigma}_n(v_N)>_{\mathcal{R}_1}}.\label{onN}
\ee
For the single interval, suppose the original CFT is defined on a complex plane $\mathcal{C}$, there is a direct conformal transformation which maps the $n$-sheeted Riemann surface $\mathcal{R}_{n,1}$ to the complex plane,
\be
\omega^n=\frac{z-u}{z-v},\label{cm}
\ee
where $z$ and $\omega$ are the coordinates defined on $\mathcal{R}_{n,1}$ and $\mathcal{C}$ respectively. Hence we find an identity due to the conformal transformation\footnote{We just label the holomorphic part.},
\be
<\mathcal{O}(z)>_{\mathcal{R}_{n,1}}=<(\omega')^{h_{\mathcal{O}}}\mathcal{O}(\omega)+\cdots>_{\mathcal{C}},\label{o}
\ee
$\omega'$ is the derivative to $z$, $h_{\mathcal{O}}$ is the conformal dimension of $\mathcal{O}$, $\cdots$ is the inhomogeneous term in the conformal transformation for a general quasi-primary operator. Let $N=1$ in (\ref{onN}) and compare it with (\ref{o}), we can solve the three point correlator $<\mathcal{O}\sigma_n\tilde{\sigma}_n>$. For example, for the stress tensor, the Schwarzian derivative will contribute to $<\mathcal{T}\sigma_n\tilde{\sigma}_n>$ and the result is consistent with those from Ward identity of $\mathcal{T}$ with two primary operator with conformal dimension $h=\frac{c}{24}(n-\frac{1}{n})$. This fact actually determines the partition function $Z_{n,1}$, hence the one-interval R$\acute{e}$nyi entropy and entanglement entropy are
\be
S_{[u,v]}^{(n)}=\frac{c(n+1)}{6n}\log\frac{|u-v|}{\epsilon},\ S_{[u,v]}=\frac{c}{3}\log\frac{|u-v|}{\epsilon}\label{oneint}.
\ee
When the system is at a finite temperature $\frac{1}{\beta}$, there is another conformal transformation which maps the complex plane to the cylinder,
\be
s=\frac{\beta}{2\pi}\log z,\label{thm}
\ee
where $s$ is a cylinder coordinate with\footnote{Here $\sigma$ is the spatial coordinate in the cylinder. One should distinguish it from the twist operator $\sigma_n$.} $s=\sigma+i\tau$, $\sigma\in(-\infty,\infty),\tau\in[0,\beta)$. Then the transformation of the twist operator under (\ref{thm}) leads to the one-interval R$\acute{e}$nyi entropy and entanglement entropy at  finite temperature,
\be
S^{(n)}_\beta([u,v])=\frac{c(n+1)}{6n}\log\frac{\beta}{\pi\epsilon}\sinh\frac{\pi|u-v|}{\beta},\ S_\beta([u,v])=\frac{c}{3}\log\frac{\beta}{\pi\epsilon}\sinh\frac{\pi|u-v|}{\beta}.\label{tee}
\ee

For one-interval, the essential part is the existence of a conformal map (\ref{cm}).  This conformal map is multi-valued, whose $j$-th solution is
\be
\omega_j(z)=\omega_0(z)\exp^{\frac{2\pi ij}{n}},\  j=0,1,\cdots,n-1,
\ee
where $\omega_0(z)=(\frac{z-u}{z-v})^{\frac{1}{n}}$.

Before we close this section, we emphasize that the conclusion (\ref{oneint}) and (\ref{tee}) rely on the partition function is\footnote{There is always a barred sector here and below. }
\be
Z_1=tr q^{L_0-\frac{c}{24}},\label{ptf}
\ee
where $q=e^{2\pi i\tau}$. Once there is another continuous global symmetry, for instance, a higher spin symmetry, the  partition function (\ref{ptf}) should be deformed to
\be
Z_1=tr q^{L_0-\frac{c}{24}}e^{-\mu\int W}.\label{ins1}
\ee
One can expand the modified partition function according to the order of the chemical potential, and solve the problem perturbatively.
There is another version of the modified partition function in the literature \cite{Kraus1111, Gaberdial1203}, which is defined by inserting a zero mode of the higher spin current to (\ref{ptf}) \be
Z_1=tr q^{L_0-\frac{c}{24}}e^{2\pi i\alpha W_0},\label{ins2}
\ee
where the parameter $\alpha=\mu\bar{\tau}$. However, the two ways of introducing the higher spin chemical potentials in (\ref{ins1}) and (\ref{ins2}) are not manifestly equivalent\footnote{We thank the anonymous referee for these helpful comments.} \cite{prob}. (\ref{ins1}) and (\ref{ins2}) may corresponds to canonical and holomorphic partition functions respectively. In this work, we will use the deformation defined by (\ref{ins1}) for our computation. Hence, one should be careful to compare our CFT result with the holomorphic result in the gravity side. However, we have checked that (\ref{ins1}) and (\ref{ins2}) are actually the same up to order $\mu^4$, including the quantum corrections, please find more details in Appendix D. Hence, at least up to order $\mu^4$, we can trust the classical and quantum results in this work. However, it is still an open issue to explain why we can obtain the same answer.

\section{AdS$_3$ Higher Spin Gravity}
The AdS$_3$ nonlinear higher spin gravity \cite{3dVasiliev} describes the interaction between scalars(and fermions when promoting supersymmetry) and an infinite tower of spins with $s\ge 1$. The matter multiplet can be truncated and one finds pure higher spin AdS$_3$ theory. This pure higher spin theory is described by two flat connection equations
\be
F=\bar{F}=0,\label{flat}
\ee
where\footnote{In this paper, we always omit the bar term to simplify notation.} $F=dA+A^2$ and the connection one form $A$ is valued in some higher spin algebra, which includes spin 2 algebra $sl(2)$ as a subalgebra. A well-known higher spin algebra is $hs[\lambda]$ which includes each spin $s\ge 2$ once. Another view is to write 3d gravity as a difference of two Chern-Simons theory \cite{Townsend, Witten},
\be
S_{gr}=S_{CS}[A]-S_{CS}[\bar{A}],\label{action}
\ee
with the Chern-Simons action $S_{CS}[A]=\frac{k_{cs}}{4\pi}\int Tr(AdA+\frac{1}{3}A^3)$, $A$ is valued in $sl(2)$. $k_{cs}$ is related to Newton constant by comparing (\ref{action}) to Einstein-Hilbert action. The pure higher spin gravity is constructed by  embedding $sl(2)$ into a larger algebra \cite{Blencowe}.
\subsection{Asymptotic Symmetry and $\mathcal{W}$ algebra}
By gauge fixing and imposing the extended asymptotic AdS$_3$ boundary condition, one finds that the asymmetric symmetry is generated by a classical $\mathcal{W}$ algebra \cite{Henneaux,Campoleoni}. Hence the CFT$_2$ dual theory have a $\mathcal{W}$ symmetry. A $\mathcal{W}$ algebra can be understood as adding some higher spin primary fields to Virasoro algebra. The quantum version of the classical $\mathcal{W}$ algebra can be found by imposing Jacobi identity condition. For the purpose of our discussion, we only give the first few OPEs of quantum $\mathcal{W}_{\infty}(\lambda)$ as follows,
\bea
\mathcal{T}(z)\mathcal{T}(0)&\sim& \frac{c/2}{z^4}+\frac{2\mathcal{T}}{z^2}+\frac{\partial{\mathcal{T}}}{z}\\
\mathcal{T}(z)\mathcal{W}(0)&\sim& \frac{3\mathcal{W}}{z^2}+\frac{\partial{\mathcal{W}}}{z}\\
\mathcal{T}(z)\mathcal{U}(0)&\sim& \frac{4\mathcal{U}}{z^2}+\frac{\partial{\mathcal{U}}}{z}\\
\frac{1}{\mathcal{N}_3}\mathcal{W}(z)\mathcal{W}(0)&\sim&\frac{1}{z^6}+\frac{\frac{6}{c}\mathcal{T}}{z^4}+\frac{\frac{3}{c}\partial\mathcal{T}}{z^3}+\frac{\frac{12}{c}\mathcal{U}+\frac{9}{10c}\partial^2\mathcal{T}+\frac{96}{c(5c+22)}\Lambda}{z^2}+\nn\\&&\frac{\frac{6}{c}\partial\mathcal{U}+\frac{1}{5c}\partial^3\mathcal{T}+\frac{48}{c(5c+22)}\partial\Lambda}{z} \\
\mathcal{W}(z)\mathcal{U}(0)&\sim&\frac{12(\lambda^2-9)}{5(\lambda^2-4)}(\frac{\mathcal{W}}{z^4}+\frac{1}{3}\frac{\partial\mathcal{W}}{z^3}+\frac{1}{14}\frac{\partial^2\mathcal{W}}{z^2}+\frac{1}{84}\frac{\partial^3\mathcal{W}}{z}+\cdots)+\cdots\\
\mathcal{U}(z)\mathcal{U}(0)&\sim&\frac{c(\lambda^2-9)}{5(\lambda^2-4)}\frac{1}{z^8}+\cdots
\eea
 Where $\mathcal{T},\mathcal{W}$ and $\mathcal{U}$ are spin 2, spin 3 and spin 4 operators correspondingly. $\Lambda$ is a composite quasi-primary operator which is defined as $\Lambda=:\mathcal{T}\mathcal{T}:-\frac{3}{10}\partial^2\mathcal{T}$.  In the last two OPEs, we just list the terms which are relevant to our discussions below. The normalization constant $\mathcal{N}_3$ can be chosen freely. When $\mathcal{N}_3=c/3$, These quantum $\mathcal{W}_{\infty}(\lambda)$ algebra is the same as \cite{Gaberdiel1205}. In the following discussion, we will choose $\mathcal{N}_3=-\frac{5c}{6\pi^2}$. There is a free parameter $\lambda$, which is related to the higher spin algebra $hs[\lambda]$ in the bulk.

In a conformal field theory, for a primary operator $\mathcal{O}_i$ with dimension $h_i$,  the two, three and four point functions are respectively
\bea
<\mathcal{O}_i(z_1)\mathcal{O}_j(z_2)>_{\mathcal{C}}&=&\delta_{ij}\frac{\mathcal{N}_{\mathcal{O}_i}}{z_{12}^{2h}},\label{2pt}\\
<\mathcal{O}_i(z_1)\mathcal{O}_j(z_2)\mathcal{O}_k(z_3)>_{\mathcal{C}}&=&\frac{C_{\mathcal{O}_i\mathcal{O}_j\mathcal{O}_k}}{z_{12}^{h_i+h_j-h_k}z_{23}^{h_j+h_k-h_i}z_{31}^{h_k+h_i-h_j}},\label{3pt}\\
<\mathcal{O}_i(z_1)\mathcal{O}_j(z_2)\mathcal{O}_k(z_3)\mathcal{O}_l(z_4)>_{\mathcal{C}}&=&\frac{(\frac{z_{24}}{z_{14}})^{h_i-h_j}(\frac{z_{13}}{z_{14}})^{h_k-h_l}}{z_{12}^{h_i+h_j}z_{34}^{h_k+h_l}}f_{ijkl}(x),\label{4pt}
\eea
with $z_{ij}=z_i-z_j$. They are determined by global conformal symmetry.   The constants $\mathcal{N}_{\mathcal{O}_i}, C_{\mathcal{O}_i\mathcal{O}_j\mathcal{O}_k}$  can be read from the $\mathcal{O}_i\mathcal{O}_j$ OPE.  And $f_{ijkl}$ is a function of the cross ratio $x=\frac{z_{13}z_{24}}{z_{14}z_{23}}$. For the theory with $\mathcal{W}_{\infty}(\lambda)$ symmetry, we find
\be
\mathcal{N}_{\mathcal{W}}=\mathcal{N}_3,\  C_{\mathcal{W}\mathcal{W}\mathcal{W}}=0
\ee
and
\be
f_{\mathcal{W}\mathcal{W}\mathcal{W}\mathcal{W}}=\mathcal{N}_3^2\sum_{i=0}^{6}a[3,j]\th^j,
\ee
where $\th(z_1,\cdots,z_4)=x+\frac{1}{x}-2=\frac{z_{12}^2z_{34}^2}{z_{13}z_{14}z_{23}z_{24}}$. The constants $a[3,j]$ are respectively
\bea
&&a[3,0]=1,\ a[3,1]=\frac{18}{c}, \ a[3,2]=-\frac{9(c-98)}{c(5c+22)}+\frac{144(\lambda^2-9)}{(\lambda^2-4)5c},\nn\\&&a[3,3]=2+\frac{54}{c},\ a[3,4]=9+\frac{18}{c},\ a[3,5]=6,\ a[3,6]=1.\label{spin3coe}
\eea
Actually, the four point correlator of spin $J$ is
\be
\frac{\mathcal{N}_J^2}{z_{12}^{2J}z_{34}^{2J}}\sum_{j=0}^{2J}a[J,j]\th^j.\label{spinj4pt}
\ee
This can be understood as follows. The exchange symmetry of 1 and 2 restrict the function $f(x)$ as a function of $\th=(x+\frac{1}{x})-2$, while the maximal power of $\th$ is determined by examining the most singular behavior as $z_1\to z_3$. Constants $a[J,j]$ are found by matching (\ref{spinj4pt}) with the spin $J$ four point function from the Ward identity.

\subsection{Higher Spin Black Hole}
A higher spin black hole \cite{kraus1103} is a solution of (\ref{flat}) which has higher spin charges and higher spin chemical potentials. Choosing the three coordinates to be $x^{\pm},\rho$, a spin 3 black hole in $sl(3)$ theory is
\be
A=b^{-1}(a+d)b,
\ee
with
\be
a=a_+dx^++a_-dx^-,\ \  b=e^{\rho L_0},
\ee
\be
a_+=L_1-\frac{2\pi}{k}\mathcal{L}L_{-1}-\frac{\pi}{2k}\mathcal{W}W_{-2},\  a_-\sim\mu(a^2_+-\frac{1}{3}tr a_+^2).
\ee
Here we split the $sl(3)$ generators into the spin 2 part($L_i,i=0,\pm1$) and the spin 3 part($W_m,m=0,\pm1,\pm2$). Two parameters $\mathcal{L},\mathcal{W}$ are related to the spin 2 and spin 3 charges\footnote{We should mention that in different formalism, the identification of the charges is different. But in any case they can be determined by these two parameters.}. The parameter $\mu$ is the spin 3 chemical potential. There is another implicit parameter in this solution, which is the inverse temperature $\tau$. It appears in the periodic identification\footnote{Here we have transfered the solution to Euclidean version and replaced $x^+(x^-)$ with $z(\bar{z})$.}
\be
z\sim z+2\pi,\  z\sim z+2\pi \tau.
\ee
To ensure the smoothness of the solution, one should impose trivial holonomy condition around the thermal circle. This condition relates the chemical potentials to the higher spin charges, hence there are two freely parameters in the smooth solutions. A spin 3 black hole in $hs[\lambda]$ theory can be constructed analogously.

 Though the usual notion of geometry is lack in this context, one can still define consistent thermodynamics for higher spin black holes. So far, several methods have been developed to study the first law thermodynamics of higher spin black holes. These includes,
\begin{enumerate}
\item Dimensional analysis \cite{Jiang1212}. In this method, one requires the consistency of first law of thermodynamics and relies on the dimensional counting of the quantities in the theory. The striking formula follows from Euler's theorem on homogeneous functions. One can regard this method as a higher spin generalization of Smarr's formula \cite{Smarr} for usual black holes.

\item Action variational principle \cite{Banados,David,Boer1302}. In this method, the action is thought as a saddle point approximation of the partition function. One adds suitable boundary terms to the action to ensure the variation of the action to be consistent with the first law of thermodynamics. It is a natural generalization of Gibbons-Hawking's analysis of black hole thermodynamics \cite{Gibbons}.
\item Conical singularity method \cite{kraus1305}. The gravitational entropy can be calculated by the conical singularity \cite{Soludukhin1104,Fursaev95} method, the authors in \cite{kraus1305} found an extension to the higher spin gravity.
\item Wilson line approach \cite{Ammon1310,Boer1306}. This method can be generalized to calculate classical holographic entanglement entropy straightforwardly, we will discuss it in more detail in the following sections.
\item Noether charge method \cite{Hijano}. After suitability reformulating the Noether charge method \cite{Wald97} in Chern-Simons language, the authors in \cite{Hijano} can define the entropy of higher spin black holes. It is interesting that another choice of the Killing parameter leads to a new entropy for a higher spin black hole.  In this paper we will not discuss this case.
\end{enumerate}
A higher spin black hole corresponds to a CFT ensemble at finite temperature and with (higher spin) chemical potentials. For a higher spin black hole with a spin 3 chemical potential turning on, assuming the chemical potential $\mu$ is small\footnote{More precisely, the dimensionless quantity $\frac{\mu}{\tau}\ll1$.} , one can evaluate the partition function and reproduce the tree level results in the gravity side\footnote{Some works on higher spin partition function can also be found in \cite{Beccaria1312,Beccaria1407}.}\cite{Kraus1111,Gaberdial1203}. Here we give the perturbative partition function for spin 3 black hole in $hs[\lambda]$ theory,
 \be
 \log Z=\frac{i\pi c}{12\tau}(1-\frac{4}{3}(\frac{\alpha}{\tau^2})^2+\frac{400(\lambda^2-7)}{27(\lambda^2-4)}(\frac{\alpha}{\tau^2})^4+\cdots)+quantum\ correction.\label{pf}
 \ee
 Here $\alpha=-\mu\tau$. For $\lambda=3$, it reproduces the partition function of $sl(3)$ black hole. The order $\mathcal{O}(c)$ part can be viewed as the tree level result and can be found either from gravity side \cite{Kraus1111}\footnote{In the gravity side,  there are canonical formalism or holomorphic formalism, the result we list above is from holomorphic formalism. But one expects the same answer after one identifies the charges and potentials properly in the canonical formalism \cite{Compere}.} or CFT side \cite{Kraus1111,Gaberdial1203}. The quantum correction begins with $\mathcal{O}(c^0)$, it should be there and there is no explicit result so far. In our perturbative approach, we will find  quantum corrections at the order $\mu^4$.
 The partition function should be \cite{Datta1406}
\be
Z=<exp{(-\mu\int \mathcal{W}})>_{\beta}.\label{pfspinj}
\ee
We have used a subscript $\beta$ to represent the thermal ensemble. One can expand the exponential according to the order of $\mu$ and calculate the partition function directly,
\be
\log Z=\log Z_0+\frac{\mu^2}{2}\int\int <\mathcal{W}\mathcal{W}>_{\beta}+\cdots
\ee
$Z_0$ is the thermal partition function without higher spin deformation.
We will develop this perturbation method to the order $\mathcal{\mu}^4$ and find the exact answer in the following section. Here ``exact''  means that we can calculate the all loop higher spin thermodynamics at a finite order $\mu^k$, including the tree level and quantum corrections to the partition function.  The tree level results (proportional to central charge $c$) will reproduce the gravity answer.

\section{Holographic HSEE}
As discussed in section 2, we can define the entanglement entropy for a QFT. Due to $AdS/CFT$ correspondence, there should be a concept which is dual to entanglement entropy in CFT. In \cite{Ryu05}, the authors introduced holographic entanglement entropy(HEE) to resolve this problem. In short, for Einstein-Hilbert theory, the static HEE is the minimal area of the surfaces which is homologous to the boundary region A, up to a coefficient $\frac{1}{4G_N}$. In $AdS_3/CFT_2$, for one-interval $[u,v]$,  this is just the length of the geodesics which connects the points $u$ and $v$. But in this case the story is more interesting. The entanglement entropy is also equal to the logarithmic of a Wilson line \cite{Boer1306,Ammon1310} which connects the two points $u$ and $v$. Since  Wilson line can be defined for arbitrary higher spin theory, it is natural to conjecture this object is the holographic HSEE. We discuss \cite{Boer1306} in detail, since it also provides the holomorphic result.

The one-interval holographic HSEE is conjectured to be
\be
S_{A}=\frac{k_{cs}}{\sigma_{1/2}}\log\lim_{\rho_0\to\infty}W_R(P,Q)|_{\rho_P=\rho_Q=\rho_0},
\ee
where $P$ and $Q$ are two bulk points. When $\rho_0\to\infty$, they tend to the points $u$ and $v$. $\sigma_{1/2}$ is a constant and can be determined by the theory. The Wilson line $W_R(P,Q)$ is defined to be
\be
W_{R}(P,Q)=tr_{R}[\mathcal{P}\exp\int_Q^P\bar{A}\ \mathcal{P}\exp\int_P^QA],\label{wl}
\ee
where $\mathcal{P}$ means path ordering and $R$ denotes the representation.  The representation can be found by matching the HEE to thermal entropy. In the holomorphic formalism, Wilson line (\ref{wl}) is replaced by
\be
W_R(P,Q)=tr_{R}[\mathcal{P}\exp\int_Q^P\bar{A}_-dx^-\mathcal{P}\exp\int_P^QA_+dx^+].\label{wlh}
\ee
For spin 3 black hole in $sl(3)$ theory, $\sigma_{1/2}=1$ and $R$ is chosen to be the adjoint representation. The HSEE in the holomorphic formalism is
\bea
S(\Delta)&=&\frac{1}{3} c \log[\frac{\beta}{\pi\epsilon} \sinh\frac{\pi\Delta}{\beta}]+
\frac{4 c \pi^2\mu^2}{9 \beta^2(-1 + U)^4}\sum_{i=0}^2 e[2,i]\log^i[U]+\nn\\&&
\frac{32 c \mu^4 \pi^4}{27 \beta^4 (-1 + U)^8}\sum_{i=0}^4e[4,i]\log^i[U]+\mathcal{O}(\mu^6),\label{thee}
\eea
where $U=e^{\frac{2\pi \Delta}{\beta}}$ and  $\Delta=|u-v|$.
We have expanded the HSEE to order $\mu^4$. The first term on the right hand side is the usual entanglement entropy (\ref{tee}). There is no contribution from odd power of $\mu$. Actually this is related to the fact that the correlation function of odd number of spin 3 operators is zero in the CFT side. For each even power of $\mu$, the $(\frac{\mu}{\beta})^{2k}$ correction is of the form
\be
\sum_{i=0}^{2k}g[2k,i;U]\log^i[U],
\ee
where $g[2k,i;U]$ is a rational function of U. In the limit $U\to\infty$,  the terms  $g[2k,1;U]\log[U]$ dominate. This is because that the HSEE should be proportional to the thermal entropy in the large interval limit,
\be
S(\Delta)\to s_{thermal}\Delta, \ as\ \Delta\to\infty.
\ee
We have rewritten $g[2,i;U],g[4,i;U]$ in (\ref{thee}) in terms of the functions $e[2,i;U]$ and $e[4,i;U]$. They are defined to be
\bea
e[2,0;U]&=& -(-1 + U)^2 (5 + 2 U + 5 U^2),\nn\\
e[2,1;U]&=&4 (-1 - U + U^3 + U^4),\\
e[2,2;U]&=& -6 (U + U^3),\nn\\
e[4,0;U]&=&
   -(-1 + U)^4 (43 - 52 U + 162 U^2 - 52 U^3 +
       43 U^4),\nn\\
e[4,1;U]&=&
    8 (-1 + U)^3 (5 + 12 U + 19 U^2 + 19 U^3 + 12 U^4 + 5 U^5),
    \nn\\
    e[4,2;U]&=&- 4 (-1 + U)^2 U (43 + 26 U + 78 U^2 + 26 U^3 + 43 U^4),\\
    e[4,3;U]&=& 12 U (-3 - 8 U + U^2 - U^4 + 8 U^5 + 3 U^6),\nn\\
    e[4,4;U]&=& -
    3 U (1 + 8 U + 7 U^2 + 16 U^3 + 7 U^4 + 8 U^5 + U^6). \nn
\eea
Before ending this section, we remark that the Wilson line conjecture only reproduce the $\mathcal{O}(c)$ part of the HSEE. It would be an interesting issue to find out the quantum correction in the bulk side.

\section{Partition Function of Higher Spin Black Hole}
The first step towards the HSEE is to find the higher spin partition function as indicated by formula (\ref{hsdf}). We  develop a different method to calculate (\ref{pf}), including its quantum correction. This is also a warmup exercise before we tackle the more tough problem on HSRE and HSEE.

Let us consider a spin $J$ chemical potential in (\ref{pfspinj}). We expand the exponential in (\ref{pfspinj}), using the fact (\ref{2pt}-\ref{4pt}) and (\ref{spinj4pt}) on the correlator of spin $J$ operator and the conformal map (\ref{thm}), then the partition function $\log Z$ can be written out order by order
\bea
\mathcal{O}(\mu^1)&:&0\\
\mathcal{O}(\mu^2)&:&\frac{\mu_J^2}{2}\mathcal{N}_J(\frac{2\pi}{\beta})^{2J}(\frac{\beta^2}{2\pi})^2I_2[J]\label{od2}\\
\mathcal{O}(\mu^3)&:&\frac{\mu_J^3}{6}C_{JJJ}(\frac{2\pi}{\beta})^{3J}(\frac{\beta^2}{2\pi})^3I_3[J]\label{od3}\\
\mathcal{O}(\mu^4)&:&\frac{\mu_J^4}{24}\mathcal{N}_J^2(\frac{2\pi}{\beta})^{4J}(\frac{\beta^2}{2\pi})^4\sum_{j=0}^{2J}a[J,j]I_4[J,j]-\frac{\mu^4}{8}\mathcal{N}_J^2(\frac{2\pi}{\beta})^{4J}(\frac{\beta}{2\pi})^4(I_2[J])^2\label{od4}
\eea
where $\mathcal{N}_J, C_{JJJ},a[J,j]$ are constants which are determined by two, three and four spin $J$ correlation functions.  Please find the details on $\mathcal{N}_J,C_{JJJ},a[J,j]$ in subsection 3.1. The integrals $I_2[J],I_3[J],I_4[J,j]$ are defined respectively as
\bea
I_2[J]&=&\int dt_1dt_2\frac{(t_1t_2)^{J-1}}{t_{12}^{2J}},\\
I_3[J]&=&\int dt_1dt_2dt_3\frac{(t_1t_2t_3)^{J-1}}{(t_{12}t_{23}t_{31})^J},\\
I_4[J,j]&=&\int dt_1dt_2dt_3dt_4\frac{(t_1t_2t_3t_4)^{J-1}}{t_{12}^{2J-2j}t_{34}^{2J-2j}t_{13}^jt_{14}^jt_{23}^jt_{24}^j}.\label{integrali4}
\eea
All the integrals are definite integrals ranging from 0 to $\infty$. Let us explain (\ref{od2}-\ref{integrali4}) in more detail. The integral  in (\ref{pfspinj}) is over the whole cylinder
\be
\int W=\int_{0}^{\beta} d\tau\int_{-\infty}^{\infty}d\sigma W.\label{integ}
 \ee
 The integral of $W$ in the spatial direction is just a conserved charge, hence we first anticipate the integral of $\tau$ to yield a factor $\beta$. To integrate $\sigma$, we use the coordinate transformation $t=e^{\frac{2\pi}{\beta}\sigma}$, hence the integral replacement rule is
 \be
 \int_{0}^{\beta}d\tau\int_{-\infty}^{\infty}d\sigma=\frac{\beta^2}{2\pi}\int_{0}^{\infty}\frac{dt}{t}.\label{it}
 \ee
 In perturbation expansion (\ref{od2}-\ref{od4}), the factor $\frac{2\pi}{\beta}$ is from the conformal map (\ref{thm}), whilst the factor $\frac{\beta^2}{2\pi}$  is from (\ref{it}). The two dimensional integrations in $I_2$ and $I_4$ become one dimensional integral ranging from 0 to $\infty$.

  However,  there are divergences originating from the infinite length of the cylinder.  We introduce an IR cutoff by setting the length of the cylinder to be L. Define $M=\frac{2\pi L}{\beta}$, then the answer can be written as a function of M.

For spin 3, the three point function is zero. We need to evaluate $I_2[3],I_4[3,j](j=0,\cdots,6)$ , they are
\bea
I_2[3]&=&-\frac{M}{30},\
I_4[3,0]=\frac{M^2}{900},\
I_4[3,1]=\frac{7 M}{648} + \frac{M^3}{1890},\nn\\
I_4[3,2]&=&\frac{M}{36},\
I_4[3,3]=-\frac{M}{36} + \frac{M^3}{630},\
I_4[3,4]=\frac{17 M}{162} - \frac{M^3}{189},\nn\\
I_4[3,5]&=&-\frac{3 M}{8} + \frac{2 M^3}{105},\
I_4[3,6]=\frac{49 M}{36} - \frac{22 M^3}{315} +\frac{M^2}{450}.\nn
\eea
Including the spin 3 coefficients (\ref{spin3coe}), we find the partition function to be
\be
\frac{2\pi}{L}\ln Z_{BH}=\frac{i\pi c}{12\tau}[1-\frac{4}{3}(\frac{\alpha}{\tau^2})^2+\frac{80  (-1154 + 206 \lambda^2 + 25 c (-7 + \lambda^2)}{27 (22 +
   5 c) (-4 + \lambda^2)}(\frac{\alpha}{\tau^2})^4+\cdots].\label{z1mu4}
\ee
The interesting fact is that the higher order divergence cancel, leaving out the linear divergence, which shows the extensive property of the partition function. If we choose the large c limit, $c\to\infty$, the $\mathcal{O}(c)$  partition function is exactly the same in (\ref{pf}). We can also read out the quantum correction at $\mathcal{O}(\mu^4)$\footnote{This is the contribution of the holomorphic part. There should be a similar anti-homomorphic part.}
\be
\log Z|_{\mu^4,quan}=\frac{i 640\pi c  }{27(22 + 5 c)\tau}(\frac{\alpha}{\tau})^4.\label{qpf}
\ee
Some remarks follow.
\begin{enumerate}
\item At the order $\mu^4$, the quantum correction to the higher spin black hole partition function  is independent of $\lambda$.
    \item In the large $c$ limit, the quantum correction contributes a finite term
\be
\frac{128 i \pi}{27\tau}(\frac{\alpha}{\tau})^4.
\ee
This is the one-loop contribution. Of course, one can expand  (\ref{qpf}) in terms of $1/c$ and find the higher loop contribution. (\ref{qpf}) contains all loop corrections to the partition function of spin 3 black hole at $\mathcal{O}(\mu^4)$.
\item  When we use another description (\ref{ins2}), we find the same partition function of higher spin black hole up to $\mathcal{O}(\mu^4)$\footnote{We thank the anonymous referee of this paper for suggesting such kind of check.}. This is included in Appendix D.
\end{enumerate}

\section{HSRE and HSEE}
In this section we compute the HSRE and HSEE to $\mathcal{O}(\mu^4)$. From (\ref{hsdf}), the unknown quantity is $\log Z_n$.

\subsection{$\mathcal{O}(\mu^2)$ Correction}
As a first step, let us  consider the $\mathcal{O}(\mu^2)$ correction. It is
\be
\frac{\mu_J^2}{2}\int ds_1ds_2<W_J(s_1)W_J(s_2)>_{\mathcal{R}_{n,\beta}}=\frac{\mu_J^2}{2}\int ds_1ds_2(\frac{2\pi}{\beta})^{2J}z_1^Jz_2^J<W_J(z_1)W_J(z_2)>_{\mathcal{R}_n}.
\ee
Here we use a spin $J$ chemical potential deformation. $\beta$ in the subscript means that we are in a thermal ensemble now. On the right hand side of the equality, we use the conformal map (\ref{thm}) from cylinder to complex plane. In this map, a point $s_i$ in the correlation function is mapped to \footnote{Note the integral variables in the right hand side are still $s_i$, the coordinates $z_i$ should be understood as $z_i=z(s_i)=e^{\frac{2\pi s_i}{\beta}}$.}$z_i=e^{\frac{2\pi s_i}{\beta}}$and the end point of the interval $u,v$ are mapped to\footnote{Since the R$\acute{e}$nyi and entanglement entropy is a function of the distance $|u-v|$, we can safely set $u=0,v=\Delta$. In this convention, the two points are mapped to $l_1=1, l_2=U$.} $l_1=e^{\frac{2\pi u}{\beta}},l_2=e^{\frac{2\pi v}{\beta}}$. To find $<W_J(z_1)W_J(z_2)>_{\mathcal{R}_n}$, we can use the conformal map (\ref{cm}) from $\mathcal{R}_n$ to complex plane. Since the theory is a $\mathbb{Z}_n$ orbifold theory, $W(z)=\sum_{j=0}^{n-1}W(z_j)$, the field in $j$-th sheet should be mapped to a value $\omega_j(z)$, hence
\be
<W_J(z_1)W_J(z_2)>_{\mathcal{R}_n}=\sum_{j_1,j_2=0}^{n-1}(\frac{\partial\omega_{j_1}}{\partial z_1})^J(\frac{\partial\omega_{j_2}}{\partial z_2})^J<W_J(\omega_{j_1})W_J(\omega_{j_2})>_{\mathcal{C}}
\ee
The summation can be found by the residue theorem,
\be
<W_J(z_1)W_J(z_2)>_{\mathcal{R}_n}=\frac{\mathcal{N}_J}{z_{12}^{2J}}\sum_{j=0}^{J-1}b[J,j;n]\xi^j,\label{2ptrn}
\ee
where $\xi$ is defined to be $\xi=\frac{z_{12}^2l_{12}^2}{(z_1-l_1)(z_2-l_1)(z_1-l_2)(z_2-l_2)}$. Please find more details on this two point correlation function on $\mathcal{R}_n$ in Appendix A. The $\mathcal{O}(\mu^2)$ correction to $\log Z_n$ is
\be
\log Z_n|_{\mu^2}=\frac{\mu_J^2}{2}\mathcal{N}_J(\frac{2\pi}{\beta})^{2J}(\frac{\beta^2}{2\pi})^2\sum_{j=0}^{J-1}b[J,j;n](U-1)^{2j}F[J,j],\label{zn2}
\ee
where the integral $F[J,j]$ is defined to be
\be
F[J,j]=\int dt_1dt_2\frac{(t_1t_2)^{J-1}}{{t_{12}^{2J-2j}(t_1-1)^j(t_1-U)^j(t_2-1)^j(t_2-U)^j}}.
\ee
Due to the identities 
\be
F[J,0]=I_2[J],\ b[J,0;n]=n,
\ee
 the $j=0$ term is cancelled in
$n$-th R$\acute{e}$nyi entropy (\ref{hsdf}), so finally
\be
S^{(n)}|_{\mu^2}=\frac{\mu_J^2}{2}\mathcal{N}_J(\frac{2\pi}{\beta})^{2J}(\frac{\beta^2}{2\pi})^2\sum_{j=1}^{J-1}\tilde{b}[J,j;n](U-1)^{2j}F[J,j].\label{regs}
\ee
 The constants $\tilde{b}[J,j;n]=\frac{b[J,j;n]}{1-n}$ whose value can be found in Appendix A. The integral $F[J,j]$ can be evaluated, please find Appendix B for details. There the reader can find the integral from spin 3 to spin 6. For spin 3 , the $\mathcal{O}(\mu^2)$ correction of R$\acute{e}$nyi entropy  is
\bea
S^{(n)}_{spin 3}|_{\mu^2}&=&-\frac{8\pi^4\mu^2\mathcal{N}_3}{\beta^2(U-1)^4}\times\nn\\&&[\frac{1 + n}{48n}(c[3,1,0]+ c[3,1,1]\log[U] +c[3,1,2]\log^2[U])+\nn\\&&
  \frac{-4 - 4 n + n^2 + n^3}{240n^3}(c[3,2,0]+c[3,2,1]\log[U]+c[3,2,2]\log^2[U])].
\eea
The definition of $c[J,j,i]$ can be found in Appendix B. Taking the limit $n\to1$, the entanglement entropy is
\be
S_{spin 3}|_{\mu^2}=-\frac{4\pi^4\mu^2\mathcal{N}_3}{15\beta^2(U-1)^4}\sum_{i=0}^2e[2,i]Log^i[U],\label{ee3}
\ee
where $e[2,i]$ is the same ones in (\ref{thee}). Once we choose normalization convention $\mathcal{N}_3=-\frac{5c}{6\pi^2}$ and take into account of the contribution from anti-holomorphic part, (\ref{ee3}) is equal to order $\mu^2$ term in (\ref{thee}).

As the function $b[J,j;n]$ can be determined by the residue theorem and $F[J,j]$ can always be integrated out, the $\mu^2_J$ correction from the spin $J$ deformation to the R$\acute{e}$nyi and entanglement entropy can be obtained. In Appendix C, we list these entropies for other spins, including the spin 4, 5 and 6 cases.

\subsection{$\mathcal{O}(\mu^3)$ Correction}
We just introduce the method briefly as the three point function of spin 3 field is vanishing. However, for other kinds of fields or when there are many chemical potentials, the three point function may not be zero. So they can contribute to the R$\acute{e}$nyi and entanglement entropy. There is only one single chemical potential $\mu_J$ in our example, but the reader can extend it to the cases with arbitrary number of chemical potentials.

As in the previous subsection, we first map finite temperature $n$-sheeted Riemann surface $\mathcal{R}_{n,\beta}$ to $\mathcal{R}_n$ and then map the $\mathcal{R}_n$ to complex plane. After carefully collecting the conformal transformation factor and noticing the multi-value of the second transformation, we find the correction is
\be
\log Z_n|_{\mu^3_J}=\frac{C_{JJJ}\mu_J^3}{6}(\frac{\pi}{\beta})^{3J}(\frac{\beta^2}{2\pi})^3\int dt_1dt_2dt_3\frac{(U-1)^{3J}}{n^{3J}}f[J,t_1]f[J,t_2]f[J,t_3]S[J,J,J;x_{12},x_{23},x_{31}].
\ee
The function $f[J,t]$ is defined as
\be
f[J,t]=\frac{t^{J-1}}{(t-1)^J(t-U)^J}
\ee
and $x_{ij}$ is
\be
x_{ij}=\frac{(t_i-1)(t_j-U)}{(t_j-U)(t_j-1)}.
\ee
The summation function $S[a,b,c;x,y,z]$ is more involved, please find it in Appendix A.
The $\mu^3$ correction depends only on the three point function of the operators. The general type of integration is
\be
\int dt R[t]\log[t]^k
\ee
with $k\le 2$ and $R[t]=\frac{P[t]}{Q[t]}$ is just a rational function of $t$ where $P[t]$ and $Q[t]$ are polynomials of $t$. All such kind of integrals can be reduced to those discussed in Appendix B. For spin 3, the three point function is just zero. At least for this special example, we need not do the summation and integral at all. But such kind of consideration is useful to compute the $\mu^4$ correction, which is indeed relevant even for spin 3.

\subsection{$\mathcal{O}(\mu^4)$ Correction}
After some algebra, we find the partition function of $n$-sheeted Riemann surface $\mathcal{R}_n$ is
\bea
\log Z_n|_{\mu_J^4}&=&\frac{\mathcal{N}_J^2\mu_J^4}{24}(\frac{\pi}{n \beta})^{4J}(\frac{\beta^2}{2\pi})^4(U-1)^{4J}\times\nn\\&&\{(\prod_{i=1}^4\int dt_if[J,t_i])\sum_{j=0}^{2J}a[J,j]S[2J-2j,j,j,j,j,2J-2j;x_{12},x_{13},x_{14},x_{23},x_{24},x_{34}]\}\nn\\
&&-\frac{1}{2}(\mathcal{O}(\mu_J^2))^2,\label{zn4}
\eea
where $\mathcal{O}(\mu_J^2)$ is (\ref{zn2}). The definition of $S$ can be found in Appendix A.3. The number of S increases linearly with spin $J$. We choose the smallest number, $J=3$ to study. In this case, there are 7 kinds of $S$. The number of independent terms is estimated of order $\mathcal{O}(10^2)$ for $j=1,2,\cdots,6$ and $\mathcal{O}(10)$ for $j=0$.  We note that the gravity result is of order c. While in R$\acute{e}$nyi entropy, there is $\mathcal{O}(c^2)$ contribution superficially. We expect this $\mathcal{O}(c^2)$ contribution to be vanish, especially when we compute the entanglement entropy. Note that one has to prove this fact from the CFT side in principal. However, we haven't proved it as the summation of the $j=6$ term is extremely difficult. Therefore we just throw out all the terms which is $\mathcal{O}(c^2)$ before computation. Next, we look for $\mathcal{O}(c)$ term, which is the combination\footnote{We choose $\mathcal{W}_3$ theory here, the $\mathcal{W}_{\infty}(\lambda)$ can be found later.}
\be
S[4,1,1,1,1,4]-\frac{1}{10}S[2,2,2,2,2,2]+3S[0,3,3,3,3,0]+S[-2,4,4,4,4,-2]\label{oc}
\ee
There are also quantum corrections, which are contributed from $j=2$ term.
\subsubsection{$\mathcal{O}(c)$}
As mentioned above, we need to evaluate $S$ for $j=1,2,3,4$.
All of the terms in S can be written  as the product of some functions like $\coth^{a_{ij}}(\log\sqrt{x_{ij}})$ with $a_{ij}$ being positive integers, $1\le i<j\le 4$. Using an identity
\be
\coth[\log{\sqrt{x_{ij}}}]=2y_{ij}-1\label{identityy}
\ee
with
\be
y_{ij}=\frac{(t_i-1)(t_j-U)}{(t_i-t_j)(1-U)}
\ee
then (\ref{oc}) becomes a function $\mathcal{P}(y_{ij})$. One can find the function $\mathcal{P}(y_{ij})$ in Appendix A.3. Each term in $\mathcal{P}(y_{ij})$ is
\be
y_{12}^ay_{i3}^by_{j4}^c
\ee
with $a,b,c$ are positive integers. Since $i<3,j<4$, there are actually six type of integrals in (\ref{zn4}). Each type of integrals can be done, please find more details in Appendix B. Finally, the answer is
\be
\log Z_n|_{\mu^4,c}=-\frac{3\mathcal{N}_3^2\mu^4\pi^8}{64c\beta^4}\frac{1}{(U-1)^8}\frac{1}{n^5}\times( \sum_{j=0}^3\sum_{i=0}^4g_{j}[i;U]n^{2j}\log^i[U]+(-\frac{16384M}{135}n^6(U-1)^8))\label{znmu4c}
\ee
with
\bea
g_3[0;U]&=&\frac{2048}{675}\times
(-12 (-1 + U)^4 (1 + 128 U + 630 U^2 + 128 U^3 + U^4))\nn\\
g_3[1;U]&=&\frac{2048}{675}\times
    8 (-1 + U)^3 (5 + 159 U + 2500 U^2 + 2500 U^3 + 159 U^4 +
       5 U^5)
       \nn\\
g_3[2;U]&=&  -\frac{2048}{675}\times
    48 (-1 + U)^2 U (8 + 277 U + 762 U^2 + 277 U^3 + 8 U^4)\nn\\
    g_3[3;U]&=& \frac{2048}{675}\times 36 U (-1 - 84 U - 421 U^2 + 421 U^4 + 84 U^5 + U^6) \nn\\
    g_3[4;U]&=& - \frac{2048}{675}\times U (1 + 224 U + 2455 U^2 + 5296 U^3 + 2455 U^4 + 224 U^5 +
        U^6)\nn
        \eea
        \bea
g_2[0;U]&=&-\frac{2048}{225}\times
    (-1 + U)^4 (5 - 2028 U - 7234 U^2 - 2028 U^3 + 5 U^4)\nn\\\
g_2[1;U]&=& -\frac{2048}{225}\times
     24 (-1 + U)^3 U (89 + 851 U + 851 U^2 + 89 U^3)\nn\\
g_2[2;U]&=&\frac{2048}{225}\times
     4 (-1 + U)^2 U (173 + 3730 U + 9114 U^2 + 3730 U^3 +
        173 U^4)\nn\\
g_2[3;U]&=&-\frac{2048}{225}\times
     4 U (-21 - 920 U - 3737 U^2 + 3737 U^4 + 920 U^5 + 21 U^6) \nn\\
g_2[4;U]&=&\frac{2048}{225}\times U (3 + 296 U + 2677 U^2 + 5328 U^3 + 2677 U^4 +
        296 U^5 + 3 U^6)\nn
\eea
\bea
g_1[0;U]&=&-\frac{2048}{225}\times
   (-1 + U)^4 (21 + 2468 U + 7070 U^2 + 2468 U^3 + 21 U^4)
   \nn\\
   g_1[1;U]&=& \frac{2048}{225}\times
     24 (-1 + U)^3 U (129 + 875 U + 875 U^2 + 129 U^3) \nn\\
     g_1[2;U]&=&-\frac{2048}{225}\times
     4 (-1 + U)^2 U (293 + 4162 U + 9162 U^2 + 4162 U^3 +
        293 U^4)\nn\\
        g_1[3;U]&=&\frac{2048}{225}\times
     8 U (-21 - 554 U - 1841 U^2 + 1841 U^4 + 554 U^5 + 21 U^6)\nn\\
     g_1[4;U]&=&-\frac{2048}{225}\times8 U (1 + 49 U + 365 U^2 + 676 U^3 + 365 U^4 + 49 U^5 +
        U^6)\nn
\eea
\bea
g_0[0;U]&=&\frac{4096}{675}\times
   3 (-1 + U)^4 (15 + 476 U + 1178 U^2 + 476 U^3 + 15 U^4) \nn\\
   g_0[1;U]&=&-\frac{4096}{675}\times
     4 (-1 + U)^3 (5 + 519 U + 2716 U^2 + 2716 U^3 + 519 U^4 +
        5 U^5) \nn\\
        g_0[2;U]&=&\frac{4096}{675}\times
     24 (-1 + U)^2 U (38 + 385 U + 774 U^2 + 385 U^3 + 38 U^4)\nn\\
     g_0[3;U]&=&-\frac{4096}{675}\times48 U (-3 - 55 U - 151 U^2 + 151 U^4 + 55 U^5 +
        3 U^6)\nn\\
        g_0[4;U]&=&\frac{4096}{675}\times
     8 U (1 + 32 U + 199 U^2 + 346 U^3 + 199 U^4 + 32 U^5 + U^6)\nn
\eea
There is a IR divergence term in $\log Z_n|_{\mu^4,c}$. It is canceled by the similar term in $n \log Z_1|_{\mu^4,c}$, so the R$\acute{e}$nyi entropy is finite at this order.  We plug (\ref{znmu4c}),(\ref{z1mu4}) into\footnote{As we are dealing with CFT with $\mathcal{W}_3$ symmetry, we choose $\lambda=3$ here.}(\ref{hsdf}) and find the $\mathcal{O}(c)$ R$\acute{e}$nyi entropy to be
\be
S^{(n)}|_{\mu^4,c}=\frac{1}{n-1}\frac{3\mathcal{N}_3^2\mu^4\pi^8}{64c\beta^4}\frac{1}{(U-1)^8}\frac{1}{n^5}\times( \sum_{j=0}^3\sum_{i=0}^4g_{j}[i;U]n^{2j}\log^i[U]).
\ee
Taking the limit $n\to1$, the entanglement entropy is
\be
S|_{\mu^4,c}=\frac{16 c \mu^4 \pi^4}{27 \beta^4 (-1 + U)^8}\sum_{i=0}^4e[4,i]\log^i[U].
\ee
Note that $e[4,i]$ are those in (\ref{thee}). After including the anti-holomorphic part, we reproduce exactly the large $c$ limit of HSEE (\ref{thee})!

\subsubsection{Quantum Correction}
As we have mentioned, there is quantum correction at $\mathcal{O}(\mu^4)$. It is from $j=2$ term in (\ref{zn4}). So it is important to derive this term seperately. The method is the same as before. We give the result of $\log Z_n|_{\mu^4,quan}$ below.
\be
\log Z_n|_{\mu^4,quan}=-\frac{\pi^8\mu^4\mathcal{N}_3^2(a[3,2])|_{quan}}{384\beta^4n^5(U-1)^8}\times[\sum_{j=0}^3\sum_{i=0}^4\gamma_{j}[i;U]n^{2j}\log^i[U]+(-\frac{1024M}{9}n^6(U-1)^8)],\label{znmu4quan}
\ee
where $(a[3,2])|_{quan}$ is the quantum parts of $a[3,2]$,
 \be
 (a[3,2])|_{quan}=\frac{4608}{5c(22+5c)}.
 \ee
 That means the quantum correction to the R$\acute{e}$nyi and entanglement entropy at $\mathcal{O}(\mu^4)$ is universal in the sense that it is independent of $\lambda$. This is similar to the quantum correction to the thermal partition function, which is also $\lambda$ independent at this order.
 There is also a IR divergence term here, it is canceled by $n \log Z_1|_{\mu^4,quan}$ exactly. So to find the R$\acute{e}$nyi entropy, we just delete the term related to M in (\ref{znmu4quan}) and then divide $1-n$. The functions $\gamma_j[i;U]$ are
\bea
\g_3[0;U]&=&\frac{256}{315}\times
  4 (-1 + U)^4 (7 - 704 U - 4150 U^2 - 704 U^3 + 7 U^4) \nn\\
  \g_3[1;U]&=&\frac{256}{315}\times
    4 (-1 + U)^3 (35 + 323 U + 10730 U^2 + 10730 U^3 + 323 U^4 +
       35 U^5)\nn\\
       \g_3[2;U]&=& -\frac{256}{315}\times
    12 (-1 + U)^2 U (37 + 2128 U + 6758 U^2 + 2128 U^3 + 37 U^4)\nn\\
    \g_3[3;U]&=&\frac{256}{315}\times4 U (-9 - 1316 U - 8429 U^2 + 8429 U^4 + 1316 U^5 +
       9 U^6)\nn\\
       \g_3[4;U]&=&  -\frac{256}{315}\times
    U (1 + 324 U + 5095 U^2 + 11336 U^3 + 5095 U^4 + 324 U^5 +
       U^6)\nn
\eea
\bea
\g_2[0;U]&=&-\frac{512}{45}\times
   2 (-1 + U)^4 (5 - 488 U - 1914 U^2 - 488 U^3 + 5 U^4)\nn\\
   \g_2[1;U]&=& -\frac{512}{45}\times
     48 (-1 + U)^3 U (19 + 221 U + 221 U^2 + 19 U^3)\nn\\
     \g_2[2;U]&=& \frac{512}{45}\times
     24 (-1 + U)^2 U (11 + 310 U + 798 U^2 + 310 U^3 + 11 U^4)\nn\\
     \g_2[3;U]&=& -\frac{512}{45}\times4 U (-7 - 440 U - 1979 U^2 + 1979 U^4 + 440 U^5 +
        7 U^6) \nn\\
        \g_2[4;U]&=&\frac{512}{45}\times
     U (1 + 132 U + 1359 U^2 + 2776 U^3 + 1359 U^4 + 132 U^5 +
        U^6)\nn
\eea
\bea
\g_1[0;U]&=&-\frac{256}{45}\times
   (-1 + U)^4 (29 + 2932 U + 8910 U^2 + 2932 U^3 + 29 U^4)\nn\\
   \g_1[1;U]&=& \frac{256}{45}\times
     24 (-1 + U)^3 U (151 + 1085 U + 1085 U^2 + 151 U^3) \nn\\
     \g_1[2;U]&=&-\frac{256}{45}\times
     12 (-1 + U)^2 U (109 + 1706 U + 3786 U^2 + 1706 U^3 +
        109 U^4) \nn\\
        \g_1[3;U]&=&\frac{256}{45}\times
     4 U (-43 - 1352 U - 4583 U^2 + 4583 U^4 + 1352 U^5 + 43 U^6)\nn\\
     \g_1[4;U]&=&-\frac{256}{45}\times U (7 + 468 U + 3585 U^2 + 6712 U^3 + 3585 U^4 +
        468 U^5 + 7 U^6)\nn
\eea
\bea
\g_0[0;U]&=&\frac{256}{315}\times(-1 + U)^4 (315 + 9676 U + 25378 U^2 + 9676 U^3 + 315 U^4)\nn\\
\g_0[1;U]&=&-\frac{256}{315}\times
     4 (-1 + U)^3 (35 + 3473 U + 19172 U^2 + 19172 U^3 + 3473 U^4 +
        35 U^5)\nn\\
        \g_0[2;U]&=& \frac{256}{315}\times
     24 (-1 + U)^2 U (246 + 2695 U + 5458 U^2 + 2695 U^3 +
        246 U^4)\nn\\
        \g_0[3;U]&=&-\frac{256}{315}\times
     16 U (-53 - 1155 U - 3201 U^2 + 3201 U^4 + 1155 U^5 +
        53 U^6) \nn\\
        \g_0[4;U]&=& \frac{256}{315}\times
     4 U (9 + 438 U + 2791 U^2 + 4864 U^3 + 2791 U^4 + 438 U^5 +
        9 U^6)\nn
\eea
The quantum correction to the entanglement entropy is
\be
S|_{\mu^4,quan}=\frac{512 c \mu^4 \pi^4}{63 \beta^4 (22 + 5 c) (-1 + U)^8}\sum_{i=0}^4q[i;U]\log^i[U],
\ee
where the functions $q[i;U]$  are respectively
\bea
q[0;U]&=&-(-1 + U)^4 (133 + 548 U + 1662 U^2 + 548 U^3 + 133 U^4)\nn\\
q[1;U]&=&4 (-1 + U)^3 (35 + 309 U + 1168 U^2 + 1168 U^3 + 309 U^4 + 35 U^5)\nn\\
q[2;U]&=&-24 ((-1 + U)^2) U (43 + 161 U + 348 U^2 + 161 U^3 + 43 U^4) \nn\\
q[3;U]&=&16 U (-11 - 91 U - 163 U^2 + 163 U^4 + 91 U^5 + 11 U^6)\nn\\
q[4;U]&=&-8 U (1 + 23 U + 97 U^2 + 136 U^3 + 97 U^4 + 23 U^5 + U^6)\nn
\eea

\subsubsection{$\mathcal{W}_{\infty}(\lambda)$ Theory}
We find the difference between $\mathcal{W}_{\infty}(\lambda)$ and $\mathcal{W}_{3}$ theory is just the $j=2$ coefficient of $a[3,j]$ at $\mathcal{O}(\mu^4)$. So the R$\acute{e}$nyi and entanglement entropy of $\mathcal{W}_{\infty}(\lambda)$ theory can be read out directly without any further computation. The difference of the R$\acute{e}$nyi entropy between $\mathcal{W}_{\infty}(\lambda)$ and $\mathcal{W}_3$ theory is
\be
S^{(n)}_{\mathcal{W}_{\infty}(\lambda)}-S^{(n)}_{\mathcal{W}_3}=\frac{1}{n-1}\frac{\pi^8\mu^4\mathcal{N}_3^2\delta}{384\beta^4n^5(U-1)^8}\times[\sum_{j=0}^3\sum_{i=0}^4\gamma_{j}[i;U]n^{2j}\log^i[U])]+\mathcal{O}(\mu^6).
\ee
Here $\delta$ is the difference of $a[3,j]$ for arbitrary $\lambda$ and $\lambda=3$, it is
\be
\delta=\frac{144 (-9 + \lambda^2)}{5 c (-4 + \lambda^2)}.
\ee
Taking the limit $n\to1$ and choosing the normalization $\mathcal{N}_3$, we find the difference of the entanglement entropy is
\be
S_{\mathcal{W}_{\infty}(\lambda)}-S_{\mathcal{W}_3}=\frac{16 c \mu^4 \pi^4(\lambda^2-9)}{63 \beta^4(\lambda^2-4)  (-1 + U)^8}\sum_{i=0}^4q[i;U]\log^i[U]+\mathcal{O}(\mu^6).
\ee
Note the difference is of $\mathcal{O}(c)$, indicating the quantum correction is the same for different $\lambda$ at this order. However, it is not clear whether this property holds to higher order of $\mu$.

\section{Conclusion and Discussion}

In this work, we have developed a perturbation formulation to calculate HSRE and HSEE at finite temperature and with finite chemical potential. As suggested in \cite{Datta1405}, by using a multi-valued conformal map from $\mathcal{R}_n$ to complex plane, the correlation functions of primary operators on $\mathcal{R}_n$ is mapped to a multi-summation of the correlation functions of the same operators in the complex plane. After doing tedious summation and integration, we reproduced the universality property which is first observed in \cite{Datta1406} and proved in \cite{Datta1405} at the order $\mathcal{O}(\mu^2)$ up to a normalization constant. This university holds  not only for spin 3 theory but also for arbitrary kinds of higher spin deformation and arbitrary number of higher spin deformations. The results for spin 4 to spin 6 are given in Appendix C, but we can extend the computation to  any spin without difficulty. We also check the holographic HSEE \cite{Ammon1310,Boer1306} up to $\mathcal{O}(\mu^4)$ for $\mathcal{W}_3$ theory. This strongly supports the Wilson line prescription of HSEE.

Besides the confirmation of the existing results in the literature, there are many novel results from our study:
\begin{enumerate}
\item We can calculate not only HSEE but also HSRE. It would be interesting to develop the dual holographic  computation of HSRE.
\item We obtained the quantum correction to HSRE and HSEE at order $\mathcal{O}(\mu^4)$ which is absent in the classical Wilson line prescription. We also find quantum correction to the partition function of higher spin black holes. These provides interesting results to probe the quantum property of higher spin gravity, for example, the fluctuation of fields on the higher spin black hole background.
\item For the higher spin black holes other than $sl(3)$ black hole, our method can also be used to calculate the HSRE and HSEE from CFT side and check the Wilson line prescription. However, as the rank of the gauge group increases, the holographic calculation becomes more difficult quickly. For the black  holes appeared in $hs[\lambda]$ theory, the exist prescription \cite{Ammon1310, Boer1306} faces technical problem due to the infinite dimension of the group. However, our CFT calculation overcome this difficulty without trouble. In fact, in this work, we have calculated the HSRE and HSEE for $\mathcal{W}_{\infty}(\lambda)$ up to $\mathcal{O}(\mu^4)$. In the holographic prescription, there is no explicit computation so far. We also found that at $\mathcal{O}(\mu^4)$, the difference of HSEE(and HSRE) between $\mathcal{W}_{\infty}$ and $\mathcal{W}_3$ theory is purely classical, without any quantum correction! 
    It would be interesting to check this fact at higher order of $\mu$.
\item There is another point we do not discuss extensively in this paper. When there are many higher spin potentials, the $\mathcal{O}(\mu^3)$ can be non-zero. Except for a theory dependent three point function constant $C_{J_1J_2J_3}$, our method indicates that the total correction of HSRE and HSEE at this order is also universal. This is similar to $\mu^2$ correction. In conformal field theory, the structure of two and three point function of primary operators are determined by global conformal symmetry, as shown in (\ref{2pt}) and (\ref{3pt}). This means the universal property found in $\mathcal{O}(\mu^2)$ and $\mathcal{O}(\mu^3)$ actually originates from the global conformal symmetry in the theory.
\end{enumerate}
Besides the previous concrete conclusion, we can give some discussions below.
\begin{enumerate}
\item We can calculate the correction to arbitrary order $\mathcal{O}(\mu^k)$ with finite\footnote{Some technical subtleties will be discussed at the end.} $k>4$. To $\mathcal{O}(\mu^k)$, we only need to know the $k$-point function of higher spin operators which is determined by the Ward identity, this may be complicated but can be evaluated.
\item Our formulation does not use the usual prescription of the twist operator. Instead, we just used a conformal map from $n$-sheeted Riemann surface to complex plane. Since the twist operator method should give the same answer as here, one may read out some interesting information about the twist operator from our result. For instance, the most singular term of the OPE between a spin $J(J>2)$ operator $\mathcal{W}_J$ with twist operator $\sigma_n$ should be
    \be
    \mathcal{W}_J(z)\sigma_n(0)\sim\frac{\sigma_n'(0)}{z^{J-1}}+\cdots
    \ee
    where $\sigma_n'$ is another primary operator with dimension $\frac{c}{24}(n-\frac{1}{n})+1$. We expect to find a complete OPE between $\mathcal{W}_J$ and $\sigma_n$, with the results found in this work.
\item When the interval number $N>1$, there is no similar conformal map. In that case, we can not get the answer for all interval lengths. However, for short intervals,we can use all the technics \cite{Calabrese0905,Calabrese1011,head1006, Bin1309} to calculate HSRE and HSEE.
\item Our method to evaluate HSEE(and HSRE) is perturbative, it give new results which is beyond the holographic method. However, we notice that the holographic description provides the classical correction to all order of $\mu$ in the large c limit. In CFT side, to get a result analogously is quite difficult due to the complicated summation and integrals  shown in the appendices. It would be interesting to develop some new methods to evaluate the all order $\mu$ result from CFT side directly.  
\end{enumerate}
There are some open issues in the computation which are listed below.
\begin{enumerate}
\item In our computation, we throw out all the $\mathcal{O}(c^2)$ terms. Moreover, for $\mathcal{O}(\mu^k)$$k>4$ correction, there are many $\mathcal{O}(c^l)$($l\ge2$) terms. It is still an open issue to prove that we can throw them out consistently.
\item When we compute the integrals, we use the replacement rules (\ref{repl1}),(\ref{repl2}),(\ref{repl3}) by hand. This leads to the correct answer. A better understanding of these rules is valuable.
\end{enumerate}

\vspace*{10mm}
\noindent {\large{\bf Acknowledgments}}\\
The author J.L thanks Zhuang-wei Jin, Feng-yan Song, Jie-qiang Wu and Jian-dong Zhang for their helpful supports during this work.  J.L is especially thankful to prof. Bin Chen for helpful discussions and for his kind proof reading and comments on previous drafts.
The work was in part supported by NSFC Grant No.~11275010.

\vspace*{5mm}

\section*{Appendix A: Correlation Function on $\mathcal{R}_n$}
\subsection*{Appendix A.1: Two Point Function on $\mathcal{R}_n$}
In (\ref{2ptrn}), we use the two point function in complex plane and notice the multi-valued solution of the conformal map (\ref{cm}), then
\be
<W_J(z_1)W_J(z_2)>_{\mathcal{R}_n}=\frac{\mathcal{N}_J}{(2n)^{2J}}(\frac{l_{12}^2}{(z_1-l_1)(z_1-l_2)(z_2-l_1)(z_2-l_2)})^J\sum_{j_1,j_2=0}^{n-1}\frac{1}{\sinh^{2J}(\frac{\log\sqrt{x}}{n}+\frac{\pi i(j_1-j_2)}{n})},\label{2ptrn2}
\ee
where the cross ratio of the four points $z_1,z_2,l_1,l_2$ is defined to be
\be
x=\frac{(z_1-l_1)(z_2-l_2)}{(z_1-l_2)(z_2-l_1)}.
\ee
(\ref{2ptrn2})is universal for different theory up to a spin $J$ normalization. This confirms the conclusion  in \cite{Datta1405}. Let's define a function
\be
S[a;x]=\sum_{j_1,j_2=0}^{n-1}\frac{1}{\sinh^{2a}(\frac{\log\sqrt{x}}{n}+\frac{\pi i(j_1-j_2)}{n})}.\label{sax}
\ee
Since the summation depends on the difference between $j_1$ and $j_2$, we get a factor $n$ by eliminating one summation index,
\be
S[a;x]=n\sum_{j=0}^{n-1}\frac{1}{\sinh^{2a}(\frac{\log\sqrt{x}}{n}+\frac{\pi ij}{n})}.
\ee
Since $a$ is an integer, the summation can be converted to a contour integral,
\be
P[a;x]=\oint_{\mathcal{C}}\frac{dz}{2\pi i}\frac{1}{\sinh^{2a}(\frac{\log\sqrt{x}}{n}+z)}\coth (nz),
\ee
in which the contour can be chosen as follows,
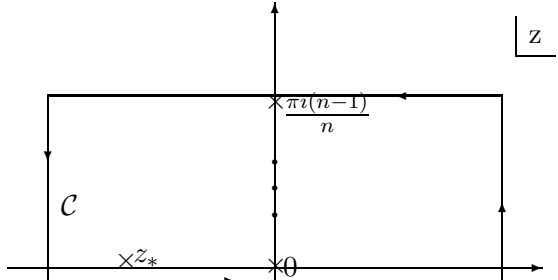
\begin{center}
\begin{picture}(120,100)
\put(-100,0){\vector(1,0){200}}
\put(0,-20){\vector(0,1){120}}
\put(-85,-5){\vector(1,0){70}}
\put(-15,-5){\line(1,0){100}}
\put(85,-5){\vector(0,1){30}}
\put(85,25){\line(0,1){40}}
\put(85,65){\vector(-1,0){40}}
\put(45,65){\line(-1,0){130}}
\put(-85,65){\vector(0,-1){25}}
\put(-85,40){\line(0,-1){45}}
\put(90,80){\line(1,0){15}}
\put(90,80){\line(0,1){15}}
\put(95,85){z}
\put(-80,20){$\mathcal{C}$}
\put(-4,-2){$\times$}
\put(-53,2){$z_*$}
\put(-4,60){$\times$}
\put(0,20){\circle*{2}}
\put(0,40){\circle*{2}}
\put(0,30){\circle*{2}}
\put(-60,0){$\times$}
\put(3,-3){$\tiny{0}$}
\put(3,55){$\frac{\pi i(n-1)}{n}$}
\end{picture}
\end{center}

The contour $\mathcal{C}$ includes the singularities $z=\frac{\pi ij}{n},j=0,\cdots,n-1$ and $z=z_*$,with $z_*=-\frac{\log\sqrt{x}}{n}$. The height of the contour is $\pi$, since under a shift of $z\to z+\pi i$ the function to be integrated is invariant when $a$ is an integer. The total contour integral is zero so that
\be
S[a;x]=-n^2 Res_{z=-\frac{\log\sqrt{x}}{n}}\frac{1}{\sinh^{2a}(\frac{\log\sqrt{x}}{n}+z)}\coth (nz).\label{residual}
\ee
We use (\ref{onN}) to organize the answer. The two point function on $\mathcal{R}_n$ should proportional to a four point function with operators $W_J,W_J,\sigma_n,\tilde{\sigma}_n$. The same reason as (\ref{spinj4pt}) tells us
\be
<W_J(z_1)W_J(z_2)>_{\mathcal{R}_n}=\frac{\mathcal{N}_J}{z_{12}^{2J}}\sum_{j=0}^{J-1}b[J,j;n]\xi^j.
\ee
Note the maximal power of $\xi$ is $J-1$, this is just a direct consequence of the residue (\ref{residual}). We evaluate the residue (\ref{residual}) for the first few spins.
 It is enough to give the $n$ related constants $b[J,j;n]$,
\bea
b[3,0;n]&=&n,\ b[3,1;n]=\frac{n^2-1}{4n},\ b[3,2;n]=\frac{n^4-5n^2+4}{120n^3}\nn\\
b[4,0;n]&=&n,\ b[4,1;n]=\frac{-1+n^2}{3 n},\  b[4,2;n]=\frac{7-10 n^2+3 n^4}{120 n^3}, \nn\\
b[4,3;n]&=&\frac{-36+49 n^2-14 n^4+n^6}{5040 n^5},\
b[5,0;n]=n,\ b[5,1;n]=\frac{5 (-1+n^2)}{ 12 n},\nn\\ b[5,2;n]&=&\frac{13-20 n^2+7 n^4}{144 n^3},\
b[5,3;n]=\frac{-164+273 n^2-126 n^4+17 n^6}{12096 n^5},\nn\\  b[5,4;n]&=&\frac{576-820 n^2+273 n^4-30 n^6+n^8}{362880 n^7},\
b[6,0;n]=n,\ b[6,1;n]=\frac{-1+n^2}{2 n},\nn\\b[6,2;n]&=&\frac{31-50 n^2+19 n^4}{240 n^3},\  b[6,3;n]=\frac{-695+1302 n^2-735 n^4+128 n^6}{30240 n^5},\nn\\b[6,4;n]&=&\frac{1916-3475 n^2+1953 n^4-425 n^6+31 n^8}{604800 n^7},\nn\\ b[6,5;n]&=&\frac{-14400+21076 n^2-7645 n^4+1023 n^6-55 n^8+n^{10}}{39916800 n^9}.\nn
\eea
The terms with $j=0$ is always $n$, because these terms should contribute to the divergent terms (proportional to M) and cancel with the terms coming from $n\log Z_1$. To simplify notation for R$\acute{e}$nyi entropy, we also introduce constants $\tilde{b}[J,j;n]=\frac{b[J,j;n]}{1-n}$ for $1\le j\le J-1$,
\bea
\tilde{b}[3,1;n]&=&\frac{-1 - n}{4 n},\ \tilde{b}[3,2;n]=\frac{4 + 4 n - n^2 - n^3}{120 n^3}\nn\\
\tilde{b}[4,1;n]&=&\frac{-1 - n}{3 n},\ \tilde{b}[4,2;n]= \frac{7 + 7 n - 3 n^2 - 3 n^3}{
 120 n^3},\ \tilde{b}[4,3;n]= -\frac{36 + 36 n - 13 n^2 - 13 n^3 + n^4 + n^5}{5040 n^5}\nn\\
 \tilde{b}[5,1;n]&=&-\frac{5 (1 + n)}{12 n},\ \tilde{b}[5,2;n]= - \frac{(1 + n) (-13 + 7 n^2)}{
 144 n^3},\ \tilde{b}[5,3;n]= - \frac{(1 + n) (164 - 109 n^2 + 17 n^4)}{
 12096 n^5},\nn\\\tilde{b}[5,4;n]&=& -\frac{(1 + n) (-576 + 244 n^2 - 29 n^4 + n^6)}{
 362880 n^7}\nn\\
 \tilde{b}[6,1;n]&=&-\frac{(1 + n)}{2 n},\ \tilde{b}[6,2;n]= - \frac{(1 + n) (-31 + 19 n^2)}{
 240 n^3},\ \tilde{b}[6,3;n]= - \frac{(1 + n) (695 - 607 n^2 + 128 n^4)}{
 30240 n^5}\nn\\ \tilde{b}[6,4;n]&=& -\frac{(1 + n) (-1916 + 1559 n^2 - 394 n^4 + 31 n^6)}{
 604800 n^7},\nn\\ \tilde{b}[6,5;n]&=& -\frac{(1 + n) (14400 - 6676 n^2 + 969 n^4 - 54 n^6 +
      n^8)}{39916800 n^9}.\nn
\eea

\subsection*{Appendix A.2: Three Point Function on $\mathcal{R}_n$}
As in the previous subsection, we map $\mathcal{R}_n$ to complex plane,
\bea
<W_{J_1}(z_1)W_{J_2}(z_2)W_{J_3}(z_3)>_{\mathcal{R}_n}&=&\frac{C_{J_1J_2J_3}}{(2n)^{J_1+J_2+J_3}}h[z_1]^{J_1}h[z_2]^{J_2}h[z_3]^{J_3}\times\nn\\&&S[J_1+J_2-J_3,J_2+J_3-J_1,J_3+J_1-J_2;r_{12},r_{23},r_{31}]
\nn\\\label{sabcx}
\eea
Here
\be
h[z]=\frac{l_{12}}{(z-l_1)(z-l_2)},\
r_{ij}=\frac{(z_i-l_1)(z_j-l_2)}{(z_i-l_2)(z_j-l_1)}
\ee
and S is a multi-summation,
\bea
S[a,b,c;x,y,z]=\sum_{j_1,j_2,j_3=0}^{n-1}&&\sinh^{-a}(\frac{\log\sqrt{x}}{n}+\frac{\pi i(j_1-j_2)}{n})\sinh^{-b}(\frac{\log\sqrt{y}}{n}+\frac{\pi i(j_2-j_3)}{n})\nn\\&&\times\sinh^{-c}(\frac{\log\sqrt{z}}{n}+\frac{\pi i(j_3-j_1)}{n})\label{sabcx2}
\eea
This summation can be done similar to (\ref{sax}). In  (\ref{sabcx}),
\be
r_{ij}r_{jk}=r_{ik}
\ee
 which is not the case for general $x,y,z$ in (\ref{sabcx2}). So for the special summation in (\ref{sabcx}),
the summation of $j_3$ is contributed by the poles at $\frac{\log\sqrt{r_{13}}}{n}+\frac{\pi ij_1}{n}$ and $\frac{\log\sqrt{r_{23}}}{n}+\frac{\pi ij_2}{n}$. This doesn't causing new poles. Then the $j_2$ summation is contributed from the poles at $\frac{\log\sqrt{r_{12}}}{n}+\frac{\pi ij_1}{n}$. The summation $j_1$ contributes a factor n as there is no pole now.

\subsection*{Appendix A.3: Four Point Function on $\mathcal{R}_n$}
We only consider the case with four higher spin operators being the same. 
\bea
<W_J(z_1)\cdots W_J(z_4)>_{\mathcal{R}_n}&=&\frac{\mathcal{N}_J^2}{(2n)^{4J}}(h[z_1]h[z_2]h[z_3]h[z_4])^J\times\nn\\&&\sum_{j=0}^{2J}a[J,j]S[2J-2j,j,j,j,j,2J-2j;r_{12},r_{13},r_{14},r_{23},r_{24},r_{34}],\nn\\
\eea
$S$ is defined as
\be
S[\{a_{ij}\};\{r_{ij}\};1\le i<j\le4]=\sum_{j_1,j_2,j_3,j_4=0}^{n-1}\prod_{1\le k<l\le4}\sinh^{-a_{kl}}(\frac{\log\sqrt{r_{kl}}}{n}+\frac{\pi i(j_k-j_l)}{n}).
\ee
This summation can be done using the residue theorem as before. In this work, we only need the four summations
\be
S[4,1,1,1,1,4],\ S[2,2,2,2,2,2],\ S[0,3,3,3,3,0],\ S[-2,4,4,4,4,-2],
\ee
each $S$ are quite lengthy. For the $\mathcal{O}(c)$ computation, we need the combination
(\ref{oc}).
As discussed in section 6.3.1, we can use the identity (\ref{identityy}) to express (\ref{oc}) as $\mathcal{P}(y_{ij})$
\bea
\mathcal{P}(y_{ij})&\equiv&-\frac{1}{n^4}(S[4,1,1,1,1,4]-\frac{1}{10} S[2,2,2,2,2,2]+3S[0,3,3,3,3,0]+S[-2,4,4,4,4,-2])\nn\\
&=&\sum_{\alpha=1}^{4}\sum_{a=1}^{2\alpha+2}n^{1+2\alpha}y_{12}^a\mathcal{\theta}_{\alpha,a}(y_{ij}).
\eea
Similarly, to compute the quantum correction, we define
\be
\mathcal{Q}(y_{ij})\equiv-\frac{1}{n^4}S[2,2,2,2,2,2]=\sum_{\alpha=1}^4\sum_{a=1}^{2\alpha+2}n^{1+2\alpha}y_{12}^a\mathcal{\psi}_{\alpha,a}(y_{ij}).
\ee
The functions $\mathcal{\theta}_{\alpha,a}$ are \footnote{To simplify notation, we define $d=y_{13},e=y_{14},f=y_{23},g=y_{24},h=y_{34}$.}
\bea
\mathcal{\th}_{1,1}&=&-\frac{8192}{675} (3 d^3 (e-h)+d (-5 (-2+e) e+3 g-6 g^2+2 (1-2 h) h)+f (e (3-6 f)+\nn\\&&e^2 (-2+4 f)+g (1-g+f (-5+3 f+3 g))+h+(1-3 f) f h+(-2+3 f) h^2)+\nn\\&&d^2 (e (-10+3 e)-2 g+4 g^2+h (2+3 h))),
\eea
\bea
\mathcal{\th}_{1,2}&=&\frac{8192}{675} (3 d^3 (e-h)-d (e (-25+9 e)+g (3+10 g)-3 h+6 h^2)+f (e^2 (2+4 f)-e (3+10 f)+\nn\\&&g (-2+3 g+3 f (f+g))-3 f^2 h+3 f h^2)+d^2 (3 (-5+e) e+2 g+4 g^2+3 h (1+h))),
\eea
\bea
\mathcal{\th}_{1,3}&=&\frac{8192}{675} (d^2 (5 e-4 g-h)+f (-4 e^2+4 e (3+f)-g (3+5 f+4 g)+h+f h-2 h^2)+\nn\\&&d (e (-21+4 e)+4 g (3+g)+h (-1+2 h))),\\
\mathcal{\th}_{1,4}&=&\frac{16384}{225} (d-f) (e-g),
\eea
\bea
\mathcal{\th}_{2,1}&=&\frac{1024}{225} (24 d^5 (e-h)+f (50 e^3 (1-2 f)+20 e^4 (-1+2 f)+4 e^2 (-1+2 f) (12+5 (-1+f) f)+\nn\\&&e (22-2 f (37+15 f (-3+2 f)))+g (5+f (-39+f (71+f (-65+24 f)))-15 g+\nn\\&&3 f (27+f (-25+9 f)) g+10 (2+f (-8+3 f)) g^2+10 (-1+3 f) g^3)+(4+f (4+\nn\\&&f (-20+(31-24 f) f))) h+3 (-6+f (4+f (-2+9 f))) h^2-10 (-3+f (2+3 f)) h^3+\nn\\&&10 (-2+3 f) h^4)+d^4 (e (-109+27 e)+20 g (-1+2 g)+h (35+27 h))+2 d (5 e (11+\nn\\&&e (-18-5 (-3+e) e))+g (11+g (-37+15 (3-2 g) g))+2 h (2+h (-9+5 (3-2 h) h)))+\nn\\&&2 d^2 (e (-105+e (114+5 e (-13+3 e)))+2 g (-1+2 g) (12+5 (-1+g) g)+h (4+\nn\\&&h (3+5 h (-1+3 h))))+d^3 (3 e (69+e (-41+10 e))-2 (25 g (-1+2 g)+h (11+\nn\\&&3 h (2+5 h))))),
\eea
\bea
\mathcal{\th}_{2,2}&=&-\frac{1024}{225} (24 d^5 (e-h)+d^4 (9 e (-17+3 e)+20 g (1+2 g)+3 h (13+9 h))-2 d (e (-400+\nn\\&&e (337+5 e (-34+9 e)))+g (11+g (131+5 g (-7+10 g)))+h (-1+2 h) (31+15 (-1+h) h))+\nn\\&&2 d^2 (e (-410+3 e (83+5 (-6+e) e))+2 g (12+g (63+5 g (-1+2 g)))+h (31+15 h (2+h^2)))+\nn\\&&d^3 (e (463+3 e (-57+10 e))-2 (5 g (5+14 g)+3 h (14+h (3+5 h))))+f (20 e^4 (1+2 f)+\nn\\&&-10 e^3 (5+14 f)+4 e^2 (12+f (63+5 f (-1+2 f)))-2 e (11+f (131+5 f (-7+10 f)))+\nn\\&&(-19+f (-35+f (55+3 f (-7+8 f)))) g+(53+3 f (19+9 (-1+f) f)) g^2+10 (-5+\nn\\&&3 (-1+f) f) g^3+30 (1+f) g^4+h (10-20 h+f (10-24 f^3+27 f^2 (1+h)+6 h (13+\nn\\&&5 (-1+h) h)-6 f (13+5 h^2))))),
\eea
\bea
\mathcal{\th}_{2,3}&=&-\frac{1024}{225} (f (-40 e^4+40 e^3 (4+f)-2 e^2 (199+4 f (34+5 f))+e (568+2 f (291+20 f (2+f)))+\nn\\&&-g (33+44 f^3+8 f^2 (13+6 g)+g (203+10 g (-1+4 g))+f (109+g (99+50 g)))+2 (12+\nn\\&&f (12+f (59+2 f))) h-2 (29+3 f (21+f)) h^2+10 (3+f) h^3-20 h^4)+d^4 (44 e-4 (10 g+h))+\nn\\&&2 d^3 (4 e (-47+6 e)+20 g (4+g)+h (61+3 h))+d (e (-2259+5 e (211-50 e+8 e^2))+\nn\\&&2 g (284+g (291+20 g (2+g)))+2 h (-1+2 h) (72+5 (-1+h) h))+d^2 (e (1321+\nn\\&&e (-393+50 e))-2 (g (199+4 g (34+5 g))+h (72+h (57+5 h))))),
\eea
\bea
\mathcal{\th}_{2,4}&=&\frac{1024}{225} (60 d^3 (-2 e+g+h)+f (60 e^3-2 e^2 (267+68 f)+2 e (742+f (313+30 f))-g (513+\nn\\&&f (377+120 f)+367 g+123 f g+60 g^2)+10 (7+f (7+6 f)) h-20 (7+3 f) h^2)+d (e (-3033+\nn\\&&(793-60 e) e)+2 g (742+g (313+30 g))+130 h (-1+2 h))-d^2 (e (-983+123 e)+\nn\\&&2 g (267+68 g)+10 h (13+6 h))),
\eea
\bea
\mathcal{\th}_{2,5}&=&-\frac{8192}{225} (d^2 (34 e-29 g-5 h)+f (-29 e^2+e (180+29 f)-g (117+34 f+29 g)+\nn\\&&5 (1+f) h-10 h^2)+d (e (-243+29 e)+g (180+29 g)+5 h (-1+2 h))),\\
\mathcal{\th}_{2,6}&=&-\frac{32768}{15} (d-f) (e-g),
\eea
\bea
\mathcal{\th}_{3,1}&=&-\frac{256}{225} (120 d^5 (e-h)+f (-4 e (-1+2 f) (17+75 (-1+f) f)+4 e^2 (-1+2 f) (53+175 (-1+f) f)+\nn\\&&-250 e^3 (-1+f (5-9 f+6 f^2))+100 e^4 (-1+f (5-9 f+6 f^2))+g (11+f (-125+f (313+\nn\\&&5 f (-65+24 f)))-61 g+3 f (121+5 f (-25+9 f)) g+50 (2+f (-8+3 f)) g^2+50 (-1+3 f) g^3)+\nn\\&&(6+f (6+f (-58+5 (31-24 f) f))) h+(-62+3 f (6+5 f (-2+9 f))) h^2-50 (-3+f (2+\nn\\&&3 f)) h^3+50 (-2+3 f) h^4)+5 d^4 (e (-109+27 e)+20 g (-1+g (5-9 g+6 g^2))+h (35+27 h))+\nn\\&&d^3 (3 e (331+5 e (-41+10 e))-250 g (-1+g (5-9 g+6 g^2))-2 h (34+15 h (2+5 h)))+\nn\\&&2 d (-5 e (-41+e (83+25 (-3+e) e))-2 g (-1+2 g) (17+75 (-1+g) g)+2 h (3+\nn\\&&h (-31+25 (3-2 h) h)))+2 d^2 (e (-455+e (549+25 e (-13+3 e)))+2 g (-1+2 g) (53+\nn\\&&175 (-1+g) g)+h (6+h (-6+25 h (-1+3 h))))),\nn\\
\eea
\bea
\mathcal{\th}_{3,2}&=&\frac{256}{225} (840 d^5 (e-h)+5 d^4 (3 e (-313+63 e)+20 g (1+g (11-3 g+6 g^2))+9 h (29+21 h))+\nn\\&&2 d^2 (e (-7045+3 e (2013+25 e (-37+7 e)))+2 g (53+g (1111+25 g (-31+38 g)))+h (134+\nn\\&&3 h (-1+5 h) (2+35 h)))+d^3 (e (11063+15 e (-351+70 e))-50 g (5+g (71+3 g (-9+\nn\\&&14 g)))-6 h (118+5 h (18+35 h)))-2 d (e (-4975+e (6557+25 e (-181+51 e)))+2 (g (17+\nn\\&&25 g (27+g (-35+38 g)))+h (-1+2 h) (67+225 (-1+h) h)))+f (100 e^4 (1+\nn\\&&f (11-3 f+6 f^2))-50 e^3 (5+f (71+3 f (-9+14 f)))-4 e (17+25 f (27+f (-35+38 f)))+\nn\\&&4 e^2 (53+f (1111+25 f (-31+38 f)))+(-67+f (-235+f (863+15 f (-93+56 f)))) g+\nn\\&&(373+9 f (117+5 f (-37+21 f))) g^2+50 (-14+3 f (-12+7 f)) g^3+150 (3+7 f) g^4+\nn\\&&h (-840 f^4+15 f^3 (67+63 h)-6 f^2 (93+5 h (3+35 h))-50 (-1+h (5-9 h+6 h^2))+\nn\\&&2 f (25+3 h (73+25 h (-6+7 h)))))),
\eea
\bea
\mathcal{\th}_{3,3}&=&-\frac{256}{225} (1440 d^5 (e-h)+f (100 e^4 (23+3 f (7+2 f))-50 e^3 (163+f (169+12 f (3+f)))+\nn\\&&2 e^2 (5902+5 f (1402+5 f (37+78 f)))-2 e (4486+f (6122+25 f (-7+166 f)))+\nn\\&&g (123+2339 g+5 (f (119+4 f (29+2 f (7+36 f)))+3 f (37+4 f (1+27 f)) g+\nn\\&&10 (-5+4 f) (11+9 f) g^2+40 (16+9 f) g^3))-2 (53+f (53+5 f (185+4 f (-37+36 f)))) h+\nn\\&&2 (281+15 f (71+f (7+54 f))) h^2-50 (21+f (43+36 f)) h^3+100 (7+18 f) h^4)+\nn\\&&20 d^4 (e (-536+81 e)+5 g (23+3 g (7+2 g))+h (124+81 h))-d (e (-66441+e (58559+\nn\\&&850 e (-35+8 e)))+2 g (4486+g (6122+25 g (-7+166 g)))+2 h (-1+2 h) (833+\nn\\&&1075 (-1+h) h))+10 d^3 (2 e (1729-597 e+90 e^2)-5 g (163+g (169+12 g (3+\nn\\&&g)))-h (235+3 h (43+60 h)))+d^2 (5 e (-12515+e (7461+10 e (-251+36 e)))+\nn\\&&2 (g (5902+5 g (1402+5 g (37+78 g)))+h (833+5 h (63+5 h (7+36 h)))))),
\eea
\bea
\mathcal{\th}_{3,4}&=&\frac{256}{225} (720 d^5 (e-h)+30 d^4 (e (-307+27 e)+50 g (3+g)+h (53+27 h))-d (e (-204081+\nn\\&&5 e (23851+30 e (-275+46 e)))+2 g (28238+5 g (3826+15 g (83+50 g)))+10 h (-1+\nn\\&&2 h) (449+195 (-1+h) h))+30 d^3 (e (1553-339 e+30 e^2)-5 g (13+2 g) (11+3 g)-h (125+\nn\\&&39 h+30 h^2))+f (1500 e^4 (3+f)-150 e^3 (13+2 f) (11+3 f)+10 e^2 (4614+f (2546+\nn\\&&15 f (39+10 f)))-2 e (28238+5 f (3826+15 f (83+50 f)))+(8781+5 f (1601+\nn\\&&6 f (193+f (133+24 f)))) g+5 (1969+3 f (371+6 f (47+9 f))) g^2+150 (15+f (29+\nn\\&&6 f)) g^3+300 (17+3 f) g^4+10 h (-131-f (131+3 f (105+f (-13+24 f)))+367 h+3 f (133+\nn\\&&3 f (7+9 f)) h-15 (21+f (13+6 f)) h^2+30 (7+3 f) h^3))+5 d^2 (e (-25499+3 e (3311+\nn\\&&10 e (-71+6 e)))+2 g (4614+g (2546+15 g (39+10 g)))+2 h (449+3 h (73+5 h (7+6 h))))),\nn\\
\eea
\bea
\mathcal{\th}_{3,5}&=&\frac{1024}{45} (12 d^4 (11 e-10 g-h)+f (-120 e^4+30 e^3 (33+5 f)-e^2 (3649+3 f (369+50 f))+\nn\\&&e (7353+f (3163+30 f (27+4 f)))-2 g (1188+f (715+6 f (53+11 f))+707 g+\nn\\&&36 f (9+2 f) g+15 (17+5 f) g^2+60 g^3)+2 (77+f (77+69 f+6 f^2)) h-2 (169+9 f (9+\nn\\&&f)) h^2+30 (3+f) h^3-60 h^4)+6 d^3 (-242 e+24 e^2+5 g (33+5 g)+h (25+3 h))+\nn\\&&d (2 e (-8433+e (3146+15 e (-43+4 e)))+g (7353+g (3163+30 g (27+4 g)))+\nn\\&&2 h (-1+2 h) (149+15 (-1+h) h))+d^2 (10 e (674+3 e (-51+5 e))-g (3649+\nn\\&&3 g (369+50 g))-2 h (149+3 h (21+5 h)))),
\eea
\bea
\mathcal{\th}_{3,6}&=&\frac{1024}{15} (4 d^3 (29 e-25 g-4 h)+d (5 e (1037+5 e (-45+4 e))-g (3169+g (827+100 g))+\nn\\&&64 (1-2 h) h)+f (-100 e^3+e^2 (881+123 f)-e (3169+f (827+100 f))+g (1657+\nn\\&&f (615+116 f)+583 g+121 f g+100 g^2)-16 (3+f (3+f)) h+16 (6+f) h^2)+\nn\\&&d^2 (e (-1205+121 e)+g (881+123 g)+16 h (4+h))),
\eea
\bea
\mathcal{\th}_{3,7}&=&\frac{4096}{15} (d^2 (65 e-61 g-4 h)+f (-61 e^2+e (504+61 f)-g (378+65 f+61 g)+\nn\\&&4 (1+f) h-8 h^2)+d (e (-630+61 e)+g (504+61 g)-4 h+8 h^2)),\\
\mathcal{\th}_{3,8}&=&\frac{172032}{5} (d-f) (e-g),
\eea
\bea
\mathcal{\th}_{4,1}&=&\frac{256}{675} (72 d^5 (e-h)+f (-36 e (-1+2 f) (1+15 (-1+f) f)+124 e^2 (-1+2 f) (1+15 (-1+f) f)+\nn\\&&-150 e^3 (-1+2 f) (1+15 (-1+f) f)+60 e^4 (-1+2 f) (1+15 (-1+f) f)+g (5+f (-67+\nn\\&&3 f (61+f (-65+24 f)))-35 g+3 f (71+3 f (-25+9 f)) g+30 (2+f (-8+3 f)) g^2+\nn\\&&30 (-1+3 f) g^3)+(2+f (2-3 f (-2+3 f) (-5+8 f))) h+(-34+3 f (2+3 f (-2+9 f))) h^2+\nn\\&&-30 (-3+f (2+3 f)) h^3+30 (-2+3 f) h^4)+3 d^4 (e (-109+27 e)+20 g (-1+2 g) (1+\nn\\&&15 (-1+g) g)+h (35+27 h))-2 d (5 e (-23+e (49+15 (-3+e) e))+18 g (-1+2 g) (1+\nn\\&&15 (-1+g) g)+2 h (-1+2 h) (1+15 (-1+h) h))+2 d^2 (e (-265+3 e (109+5 e (-13+3 e)))+\nn\\&&62 g (-1+2 g) (1+15 (-1+g) g)+h (-1+3 h) (-2+15 h^2))+3 d^3 (e (197+3 e (-41+\nn\\&&10 e))-50 g (-1+2 g) (1+15 (-1+g) g)-6 h (2+h (2+5 h)))),
\eea
\bea
\mathcal{\th}_{4,2}&=&-\frac{256}{675} (2232 d^5 (e-h)+3 d^4 (-4083 e+837 e^2+20 g (1+g (47+15 g (-13+14 g)))+\nn\\&&3 h (383+279 h))-2 d (e (-10525+3 e (5257+5 e (-769+219 e)))+2 (g (9+g (1319+\nn\\&&15 g (-341+350 g)))+39 h (-1+2 h) (1+15 (-1+h) h)))+2 d^2 (3 e (-5485+e (5059+\nn\\&&15 e (-161+31 e)))+2 g (31+g (2609+15 g (-691+722 g)))+3 h (26+5 h (-22+\nn\\&&3 h (-5+31 h))))+3 d^3 (3 e (3081+e (-1527+310 e))-2 (5 g (5+g (299+15 g (-81+86 g)))+\nn\\&&h (202+3 h (78+155 h))))+f (60 e^4 (1+f (47+15 f (-13+14 f)))-30 e^3 (5+f (299+\nn\\&&15 f (-81+86 f)))-4 e (9+f (1319+15 f (-341+350 f)))+4 e^2 (31+f (2609+\nn\\&&15 f (-691+722 f)))+(-37+3 f (-95+3 f (225+f (-437+248 f)))) g+3 (193+\nn\\&&f (857+9 f (-173+93 f))) g^2+30 (-50+3 f (-56+31 f)) g^3+90 (11+31 f) g^4+\nn\\&&3 h (-744 f^4+f^3 (897+837 h)-2 f^2 (139+15 h (3+31 h))+10 (1+h (-17+15 (3-2 h) h))+\nn\\&&2 f (5+h (79+15 h (-26+31 h)))))),
\eea
\bea
\mathcal{\th}_{4,3}&=&\frac{256}{675} (12960 d^5 (e-h)+f (60 e^4 (107+15 f (7+2 f+24 f^2))-30 e^3 (751+f (829+\nn\\&&60 f (-3+49 f)))-2 e (10242+f (14938+15 f (-1147+3694 f)))+2 e^2 (15382+\nn\\&&3 f (6466+5 f (-911+4710 f)))+g (69+f (637+12 f (-49+4 f (-151+270 f)))+\nn\\&&4709 g+9 f (53+4 f (-281+405 f)) g+30 (-631+5 f (-77+108 f)) g^2+120 (166+\nn\\&&135 f) g^3)-2 (31+f (31+3 f (689+16 f (-148+135 f)))) h+2 (527+9 f (271+\nn\\&&f (31+810 f))) h^2-30 (93+f (571+540 f)) h^3+60 (31+270 f) h^4)+12 d^4 (e (-7226+\nn\\&&1215 e)+5 g (107+15 g (7+2 g+24 g^2))+h (1786+1215 h))+6 d^3 (2 e (20419+\nn\\&&3 e (-2689+450 e))-5 g (751+g (829+60 g (-3+49 g)))-h (1291+3 h (571+\nn\\&&900 h)))-d (e (-331647+5 e (73453+6 e (-7195+1744 e)))+2 (g (10242+g (14938+\nn\\&&15 g (-1147+3694 g)))+571 h (-1+2 h) (1+15 (-1+h) h)))+d^2 (e (-376153+\nn\\&&3 e (88653+50 e (-679+108 e)))+2 (g (15382+3 g (6466+5 g (-911+4710 g)))+\nn\\&&h (571+3 h (-993+5 h (31+540 h)))))),
\eea
\bea
\mathcal{\th}_{4,4}&=&-\frac{256}{225} (9360 d^5 (e-h)+30 d^4 (e (-2671+351 e)+10 g (51+g (41+12 g (3+g)))+\nn\\&&h (569+351 h))+f (300 e^4 (51+f (41+12 f (3+f)))-30 e^3 (2147+5 f (349+42 f (7+\nn\\&&4 f)))+2 e^2 (56022+f (46478+75 f (471+442 f)))-2 e (51762+f (44338+15 f (1847+\nn\\&&2890 f)))+(6793+f (6497+30 f (189+f (409+312 f)))) g+(8377+3 f (1691+\nn\\&&30 f (131+117 f))) g^2+30 (-173+5 f (77+78 f)) g^3+300 (101+39 f) g^4+\nn\\&&10 h (-31-f (31+3 f (73+f (-289+312 f)))+527 h+3 f (197+93 f+351 f^2) h-15 (93+\nn\\&&f (109+78 f)) h^2+30 (31+39 f) h^3))+30 d^3 (e (9709+3 e (-989+130 e))-g (2147+\nn\\&&5 g (349+42 g (7+4 g)))-3 h (71+h (109+130 h)))-d (e (-698533+e (577243+\nn\\&&30 e (-8927+1790 e)))+2 (g (51762+g (44338+15 g (1847+2890 g)))+545 h (-1+\nn\\&&2 h) (1+15 (-1+h) h)))+d^2 (e (-584723+3 e (104591+50 e (-623+78 e)))+\nn\\&&2 (g (56022+g (46478+75 g (471+442 g)))+5 h (109+3 h (-223+5 h (31+78 h)))))),
\eea
\bea
\mathcal{\th}_{4,5}&=&\frac{1024}{225} (2160 d^5 (e-h)+30 d^4 (e (-877+81 e)+20 g (16+3 g (3+g))+h (155+81 h))+\nn\\&&-d (2 e (-288783+e (176264+75 e (-829+130 e)))+g (148725+g (98743+150 g (375+\nn\\&&172 g)))+370 h (-1+2 h) (1+15 (-1+h) h))+f (600 e^4 (16+3 f (3+f))-150 e^3 (339+\nn\\&&f (203+6 f (15+2 f)))+e^2 (116293+75 f (981+2 f (257+72 f)))-e (148725+\nn\\&&f (98743+150 f (375+172 f)))+2 (g (12087+f (10054+15 f (553+f (355+72 f)))+\nn\\&&10034 g+45 f (183+f (125+27 f)) g+75 (101+f (77+18 f)) g^2+150 (47+9 f) g^3)+\nn\\&&-5 (19+f (19+3 f (5+f (-43+72 f)))) h+5 (323+9 f (27+f (19+27 f))) h^2-75 (57+\nn\\&&f (37+18 f)) h^3+150 (19+9 f) h^4))+30 d^3 (e (4361-969 e+90 e^2)-5 g (339+\nn\\&&g (203+6 g (15+2 g)))-h (61+3 h (37+30 h)))+d^2 (2 e (-177364+75 e (933+\nn\\&&e (-203+18 e)))+g (116293+75 g (981+2 g (257+72 g)))+10 h (37+\nn\\&&3 h (-87+95 h+90 h^2)))),
\eea
\bea
\mathcal{\th}_{4,6}&=&-\frac{1024}{15} (48 d^5 (e-h)+f (60 e^4 (11+3 f)-100 e^3 (49+3 f (6+f))+e^2 (15749+\nn\\&&3 f (2309+60 f (11+f)))-e (28589+f (14183+20 f (266+51 f)))+g (8303+\nn\\&&f (5277+2 f (1351+3 f (125+8 f)))+5245 g+f (2767+54 f (15+f)) g+\nn\\&&20 (137+3 f (14+f)) g^2+60 (13+f) g^3)-2 (6+f (6+f (-10+3 f (3+8 f)))) h+\nn\\&&2 (102+f (62+27 f (2+f))) h^2-60 (9+f (4+f)) h^3+60 (6+f) h^4)+6 d^4 (3 e (-61+3 e)+\nn\\&&10 g (11+3 g)+h (25+9 h))-d (e (-74453+25 e (1305+4 e (-82+9 e)))+g (28589+\nn\\&&g (14183+20 g (266+51 g)))+16 h (-1+2 h) (1+15 (-1+h) h))+2 d^3 (e (4207-603 e+\nn\\&&30 e^2)-50 g (49+3 g (6+g))-2 h (16+3 h (12+5 h)))+d^2 (e (-32705+e (8941+\nn\\&&60 (-21+e) e))+g (15749+3 g (2309+60 g (11+g)))+4 h (4+h (-32+15 h (3+h))))),\nn\\
\eea
\bea
\mathcal{\th}_{4,7}&=&-\frac{2048}{15} (12 d^4 (11 e-10 g-h)+6 d^3 (12 e (-29+2 e)+5 g (55+9 g)+h+3 h^2)+\nn\\&&f (-120 e^4+30 e^3 (55+9 f)-2 e^2 (4219+27 f (42+5 f))+2 e (11466+f (3967+\nn\\&&15 f (57+4 f)))-g (10143+132 f^3+24 f^2 (53+6 g)+g (4469+30 g (43+4 g))+\nn\\&&f (4477+3 g (441+50 g)))+2 (1+f-3 f^2+6 f^3) h-2 (17+9 f (1+f)) h^2+30 (3+\nn\\&&f) h^3-60 h^4)+d (e (-43785+e (13037+30 e (-69+4 e)))+2 g (11466+g (3967+\nn\\&&15 g (57+4 g)))+2 h (-1+2 h) (1+15 (-1+h) h))+d^2 (5 e (2609-441 e+\nn\\&&30 e^2)-2 (g (4219+27 g (42+5 g))+h-9 h^2+15 h^3))),
\eea
\bea
\mathcal{\th}_{4,8}&=&\frac{14336}{5} (20 d^3 (-e+g)-3 d^2 (e (-89+7 e)+g (73+9 g))+f (20 e^3-3 e^2 (73+9 f)+\nn\\&&e (993+f (213+20 f))-g (609+20 f^2+5 g (33+4 g)+3 f (55+7 g)))+d (e (-1473+\nn\\&&(267-20 e) e)+g (993+g (213+20 g)))),
\eea
\bea
\mathcal{\th}_{4,9}&=&\frac{688128}{5} (d^2 (-e+g)-d ((-12+e) e+g (10+g))+f (e^2-e (10+f)+g (8+f+g))),\nn\\\\
\mathcal{\th}_{4,10}&=&-\frac{1376256}{5}(d-f) (e-g).
\eea

The functions $\psi_{\alpha,a}$ are
\bea
\psi_{1,1}&=&\frac{256}{315} (52 d^3 (-e+h)+d (70 (-2+e) e+27 g (-1+2 g)+43 h (-1+2 h))-d^2 (e (-165+\nn\\&&52 e)+18 g (-1+2 g)+h (43+52 h))-f (e (27-54 f)+18 e^2 (-1+2 f)+(34+\nn\\&&f (-95+52 f)) g+(-34+52 f) g^2+h (9-18 h+f (9-52 f+52 h)))),
\eea
\bea
\psi_{1,2}&=&\frac{256}{315} (52 d^3 (e-h)+f (18 e^2 (1+2 f)-9 e (3+10 f)+g (2 (-9+g)+f (-25+52 f+\nn\\&&52 g))-(25+f (25+52 f)) h+2 (25+26 f) h^2)-d (2 e (-150+53 e)+9 g (3+10 g)+\nn\\&&77 h (-1+2 h))+d^2 (e (-235+52 e)+18 g (1+2 g)+h (77+52 h))),
\eea
\bea
\psi_{1,3}&=&\frac{512}{315} (d^2 (35 e-18 g-17 h)+f (-18 e^2+18 e (3+f)-g (1+35 f+18 g)+17 (1+\nn\\&&f) h-34 h^2)+d (e (-107+18 e)+18 g (3+g)+17 h (-1+2 h))),\\
\psi_{1,4}&=&\frac{1536}{35} (d-f) (e-g),
\eea
\bea
\psi_{2,1}&=&\frac{64}{45} (192 d^5 (e-h)+3 d (-30 (-2+e) e+7 (1-2 g) g+23 (1-2 h) h)+96 d^4 ((-7+e) e+\nn\\&&h (5+h))-4 d^3 (e (-219+56 e)+h (99+64 h))+d^2 (3 e (-185+68 e)+14 g (-1+2 g)+\nn\\&&h (69+268 h))+f (e (21-42 f)+14 e^2 (-1+2 f)+(30+f (-177+4 f (107+24 f (-5+\nn\\&&2 f)))) g+2 (-15+2 f (27+8 f (-5+3 f))) g^2+h (7-14 h+\nn\\&&f (7+76 h+4 f (-35-32 h+24 f (3-2 f+h)))))),
\eea
\bea
\psi_{2,2}&=&\frac{64}{45} (192 d^5 (-e+h)+f (-14 e^2 (1+14 f)+7 e (3+58 f)+g (-18+66 g+f (255-276 g+\nn\\&&4 f (-85+24 g-24 f (-3+2 f+g))))+(55+f (55+4 f (61+24 f (-1+2 f)))) h+\nn\\&&-2 (55+122 f+48 f^3) h^2)-96 d^4 ((-9+e) e+h (7+h))+d (6 e (-250+97 e)+\nn\\&&7 g (3+58 g)+491 h (-1+2 h))+4 d^3 (e (-421+72 e)+h (253+96 h))-d^2 (3 e (-695+\nn\\&&188 e)+14 g (1+14 g)+h (491+820 h))),
\eea
\bea
\psi_{2,3}&=&\frac{128}{45} (96 d^4 (-e+h)+f (14 e^2 (13+12 f)-14 e (33+37 f)+g (-33+102 g+f (8 f (17+\nn\\&&12 f+4 g)+27 (3+8 g)))-(211+f (211+8 f (29+12 f))) h+2 (211+4 f (33+8 f)) h^2)+\nn\\&&-d (3 e (-727+210 e)+14 g (33+37 g)+571 h (-1+2 h))-8 d^3 (e (-73+4 e)+h (61+8 h))+\nn\\&&d^2 (3 e (-563+104 e)+14 g (13+12 g)+h (571+456 h))),
\eea
\bea
\psi_{2,4}&=&\frac{128}{15} (60 d^3 (-e+h)+f (-28 e^2 (4+f)+7 e (61+24 f)-g (99+96 g+4 f (39+15 f+11 g))+\nn\\&&60 (2+f (2+f)) h-60 (4+f) h^2)-4 d^2 (e (-106+11 e)+7 g (4+g)+15 h (3+h))+\nn\\&&d (e (-959+184 e)+7 g (61+24 g)+180 h (-1+2 h))),
\eea
\bea
\psi_{2,5}&=&\frac{512}{15} (d^2 (-29 e+14 g+15 h)+f (14 e^2-7 e (15+2 f)+g (62+29 f+14 g)-15 (1+f) h+\nn\\&&30 h^2)-d (2 e (-74+7 e)+7 g (15+2 g)+15 h (-1+2 h))),\\
\psi_{2,6}&=&-\frac{3584}{3} (d-f) (e-g),
\eea
\bea
\psi_{3,1}&=&\frac{128}{45} (120 d^5 (-e+h)+f (e^2 (2-4 f)+e (-3+6 f)+g (6 (-1+g)+f (75-48 g+\nn\\&&-4 f (62-25 g+15 f (-5+2 f+g))))+(-1+f (-1+4 f (17+15 f (-3+2 f)))) h+\nn\\&&2 (1-2 f (7+5 f (-4+3 f))) h^2)-60 d^4 ((-7+e) e+h (5+h))+3 d (10 (-2+e) e+\nn\\&&g (-1+2 g)+9 h (-1+2 h))+4 d^3 (e (-132+35 e)+h (57+40 h))-d^2 (3 e (-95+\nn\\&&36 e)-2 g+4 g^2+h (27+148 h))),
\eea
\bea
\psi_{3,2}&=&\frac{128}{45} (840 d^5 (e-h)-d (6 e (-250+99 e)+g (3+250 g)+653 h (-1+2 h))+60 d^4 (e (-57+\nn\\&&7 e)+h (43+7 h))-4 d^3 (e (-1312+285 e)+h (667+360 h))+d^2 (27 e (-145+44 e)+\nn\\&&2 g (1+62 g)+h (653+1948 h))+f (2 e^2 (1+62 f)-e (3+250 f)+(18+f (-285+\nn\\&&4 f (262+15 f (-27+14 f)))) g+6 (-7+2 f (24+5 f (-9+7 f))) g^2+h (-25+50 h+\nn\\&&f (-25+148 h-4 f (67+15 f (-13+14 f-7 h)+60 h))))),
\eea
\bea
\psi_{3,3}&=&\frac{256}{45} (720 d^5 (-e+h)+20 d^3 (e (-367+61 e)+220 h+86 h^2)+f (-2 e^2 (121+180 f)+\nn\\&&e (606+962 f)+g (21-72 g-5 f (-3+78 g+4 f (17-11 g+3 f (-11+12 f+6 g))))+\nn\\&&(313+f (313+20 f (20+3 f+36 f^2))) h-2 (313+10 f (27+2 f (7+9 f))) h^2)+\nn\\&&d (3 e (-1731+544 e)+606 g+962 g^2+1993 h (-1+2 h))-60 d^4 (e (-61+6 e)+\nn\\&&h (49+6 h))-d^2 (15 e (-535+126 e)+242 g+360 g^2+h (1993+3540 h))),
\eea
\bea
\psi_{3,4}&=&\frac{256}{15}(120 d^5 (e-h)+f (20 e^2 (22+13 f)-e (1441+960 f)+g (9 (19+30 g)+10 f (33+\nn\\&&29 g+2 f (7+4 g+3 f (4+2 f+g))))-20 (28+f (28+f (25+6 f (3+f)))) h+20 (56+\nn\\&&f (32+f (14+3 f))) h^2)+60 d^4 ((-16+e) e+h (14+h))-d (e (-5471+1330 e)+\nn\\&&g (1441+960 g)+1740 h (-1+2 h))-20 d^3 (e (-147+16 e)+h (105+26 h))+\nn\\&&10 d^2 (e (-517+89 e)+2 (g (22+13 g)+h (87+92 h)))),
\eea
\bea
\psi_{3,5}&=&\frac{1024}{3} (12 d^4 (e-h)+f (-4 e^2 (11+3 f)+2 e (99+34 f)-g (61+34 g+f (55+18 g+\nn\\&&2 f (13+6 f+2 g)))+(59+f (59+2 f (19+6 f))) h-2 (59+f (21+4 f)) h^2)+\nn\\&&d (e (-461+82 e)+2 g (99+34 g)+113 h (-1+2 h))+d^3 (-82 e+4 e^2+2 h (35+4 h))+\nn\\&&-d^2 (5 e (-53+6 e)+4 g (11+3 g)+h (113+66 h))),
\eea
\bea
\psi_{3,6}&=&\frac{2048}{3} (9 d^3 (e-h)+f (e^2 (19+2 f)-e (131+23 f)+g (17 (4+g)+f (35+9 f+4 g))+\nn\\&&-9 (3+f (3+f)) h+9 (6+f) h^2)+d^2 (-70 e+4 e^2+g (19+2 g)+9 h (4+h))-d (5 e (-43+\nn\\&&5 e)+g (131+23 g)+36 h (-1+2 h))),
\eea
\bea
\psi_{3,7}&=&2048 (d^2 (5 e-2 g-3 h)+f (-2 e^2+2 e (14+f)-g (21+5 f+2 g)+3 (1+f) h-6 h^2)+\nn\\&&d (e (-35+2 e)+2 g (14+g)-3 h+6 h^2)),\\
\psi_{3,8}&=&14336 (d-f) (e-g),
\eea
\bea
\psi_{4,1}&=&\frac{64}{315} (336 d^5 (e-h)+d (-70 (-2+e) e+3 g-6 g^2+67 (1-2 h) h)+f (e (3-6 f)+\nn\\&&e^2 (-2+4 f)+g (-10 (-1+g)+f (-191+124 g+4 f (171-70 g+42 f (-5+2 f+g))))+\nn\\&&h+f (1-12 f (15+14 f (-3+2 f))) h+2 (-1+2 f (17+14 f (-4+3 f))) h^2)+\nn\\&&168 d^4 ((-7+e) e+h (5+h))-4 d^3 (e (-367+98 e)+h (157+112 h))+d^2 (e (-765+\nn\\&&292 e)-2 g+4 g^2+h (67+404 h))),
\eea
\bea
\psi_{4,2}&=&\frac{64}{315} (-10416 d^5 (e-h)+d (2 e (-5850+2333 e)+g (3+1018 g)+6013 h (-1+2 h))\nn\\&&-168 d^4 (e (-249+31 e)+h (187+31 h))+4 d^3 (e (-15473+3486 e)+h (7619+4368 h))+\nn\\&&-d^2 (e (-41155+12892 e)+2 g+508 g^2+h (6013+21740 h))+f (-2 e^2 (1+254 f)+\nn\\&&e (3+1018 f)+(-54+f (2305-4 f (3125+42 f (-123+62 f)))) g+(118-4 f (577+\nn\\&&42 f (-41+31 f))) g^2+h (65-130 h+f (65-572 h+4 f (563+42 f (-61+\nn\\&&62 f-31 h)+840 h))))),
\eea
\bea
\psi_{4,3}&=&\frac{128}{315} (30240 d^5 (e-h)+f (e^2 (2186+3864 f)-2 e (2733+4957 f)+g (-59+366 g+\nn\\&&7 f (-91+564 g+8 f (248-239 g+3 f (-239+180 f+90 g))))-(2973+f (2973+\nn\\&&56 f (71+3 f (-59+180 f)))) h+2 (2973+56 f (51+f (31+135 f))) h^2)-d (e (-88103+\nn\\&&28590 e)+5466 g+9914 g^2+44133 h (-1+2 h))+168 d^4 (e (-841+90 e)+h (661+90 h))+\nn\\&&-56 d^3 (e (-4462+841 e)+h (2479+1142 h))+d^2 (7 e (-30221+7788 e)+2186 g+\nn\\&&3864 g^2+3 h (14711+35616 h))),
\eea
\bea
\psi_{4,4}&=&\frac{128}{105} (-21840 d^5 (e-h)+f (-84 e^2 (104+81 f)+e (27301+22344 f)+g (-3 (579+1036 g)+\nn\\&&28 f (-2 (63+67 g)+5 f (13+16 g-6 f (-8+26 f+13 g))))+140 (98+f (98+f (95+\nn\\&&12 f (9+13 f)))) h-140 (196+f (157+2 f (62+39 f))) h^2)+d (e (-180797+47572 e)+\nn\\&&g (27301+22344 g)+83580 h (-1+2 h))-840 d^4 (e (-148+13 e)+h (122+13 h))+\nn\\&&140 d^3 (e (-1947+296 e)+h (1215+436 h))-28 d^2 (e (-10759+2234 e)+3 g (104+81 g)+\nn\\&&5 h (597+997 h))),
\eea
\bea
\psi_{4,5}&=&\frac{512}{15} (720 d^5 (e-h)+d^2 (e (-24221+3900 e)+6 g (231+100 g)+8555 h+9780 h^2)+\nn\\&&f (6 e^2 (231+100 f)-3 e (1825+862 f)+g (908+686 g+f (881+540 g+20 f (19+20 g+\nn\\&&6 f (10+6 f+3 g))))-5 (499+f (499+4 f (115+12 f (8+3 f)))) h+10 (499+\nn\\&&2 f (153+2 f (38+9 f))) h^2)-d (-21022 e+4406 e^2+3 g (1825+862 g)+8555 h (-1+2 h))+\nn\\&&120 d^4 (e (-46+3 e)+h (40+3 h))-20 d^3 (e (-803+92 e)+h (563+148 h))),
\eea
\bea
\psi_{4,6}&=&\frac{512}{3} (48 d^5 (-e+h)+f (-4 e^2 (151+38 f)+e (3109+908 f)+g (-913-380 g-4 f (173+\nn\\&&58 g+3 f (41+10 g+2 f (15+2 f+g))))+12 (101+f (101+f (75+34 f+4 f^2))) h+\nn\\&&-12 (202+f (87+2 f (12+f))) h^2)-24 d^4 ((-27+e) e+h (25+h))+d (e (-8053+1300 e)+\nn\\&&g (3109+908 g)+2736 h (-1+2 h))-4 d^2 (2 e (-785+92 e)+g (151+38 g)+\nn\\&&171 h (4+3 h))+12 d^3 (3 e (-79+6 e)+h (187+32 h))),
\eea
\bea
\psi_{4,7}&=&-2048 (12 d^4 (e-h)+f (-2 e^2 (29+4 f)+e (412+74 f)-g (183+44 g+f (107+\nn\\&&18 g+4 f (13+3 f+g)))+(127+f (127+4 f (16+3 f))) h-2 (127+34 f+4 f^2) h^2)+\nn\\&&4 d^3 ((-27+e) e+2 h (12+h))+d (e (-785+92 e)+412 g+74 g^2+207 h (-1+2 h))+\nn\\&&-d^2 (5 e (-79+6 e)+58 g+8 g^2+23 h (9+4 h))),
\eea
\bea
\psi_{4,8}&=&-2048 (20 d^3 (e-h)+f (2 e^2 (17+f)-e (373+38 f)+g (229+90 f+20 f^2+\nn\\&&6 (5+f) g)-20 (2+f)^2 h+20 (8+f) h^2)+2 d^2 (3 (-27+e) e+g (17+g)+10 h (5+h))+\nn\\&&-d (7 e (-79+6 e)+g (373+38 g)+100 h (-1+2 h))),
\eea
\bea
\psi_{4,9}&=&-8192 (d^2 (7 e-2 g-5 h)+f (-2 e^2+e (45+2 f)-g (36+7 f+2 g)+5 (1+f) h-10 h^2)+\nn\\&&d (2 (-27+e) e+g (45+2 g)+5 h (-1+2 h))),\\
\psi_{4,10}&=&-73728 (d-f) (e-g).
\eea

\subsection*{Appendix A.4: Higher Point Function on $\mathcal{R}_n$}
As the $m(m\ge5)$  point function in complex plane can be solved by using Ward identities recursively. The m point function on $\mathcal{R}_n$ can be transformed to a set of summations. The basic k summation function is
\be
S[\{a_{ij}\};\{r_{ij}\};1\le i<j\le m]=\sum_{j_1,\cdots,j_m=0}^{n-1}\prod_{1\le k<l\le m}\sinh^{-a_{kl}}(\frac{\log\sqrt{r_{kl}}}{n}+\frac{\pi i(j_k-j_l)}{n}).
\ee
This function is reduced to two,three and four point case when $m=2,3,4$. Though residue theorem can be used to compute it, it would become tedious quickly as m increases. As an example, for four spin 3 point function on $\mathcal{R}_n$, the number of independent terms is expected to be $\mathcal{O}(10^2)$ as it is proportional to six point functions(four spin 3 and two twist operator). So for $m>4$, to find an explicit answer is horrible.  It is interesting to study this summation in other methods, this would be important to read out the information of $\mu^m$ correction to R$\acute{e}$nyi and entanglement entropy.

\section*{Appendix B: Integrals}
All the integrals appeared in this work can be converted to a few types of basic integral labeled by two real number $p$ and q,
\be
G[p,q;x]=\int^xdx\log^p x/(x-a)^q.
\ee
In general, p and q can be real numbers. However, in this work p is always a non-negative integer and q is an integer. If $q\le0$, after doing integral by parts, we can find the answer. Also, if $p=0$, the answer is simple. So we only focus on positive integers $p,q$.  Due to the recursion relation
\be
\frac{1}{(x-a)^q}=\frac{1}{q-1}\frac{d\frac{1}{(x-a)^{q-1}}}{da},
\ee
it is safe to study $q=1$. Actually to calculate the $\mathcal{O}(\mu^k)$ correction of HSEE, we only need the integrals $p\le k-1$. In this work, $k\le4$, then there are three types of integrals.
\begin{enumerate}
\item $p=1$.
\be
G[1,1;x]=\log[x] \log[1 - x/a] + PolyLog[2, x/a]
\ee
The actual integral is from $0 \to \infty$, hence the limit behaviour $x\to0$ and $x\to\infty$ of $G[p,q;x]$ is important.
\bea
\lim_{x\to0}G[1,1;x]&=&0\\
\lim_{x\to\infty}G[1,1;x]&=&\frac{1}{6} (-\pi^2 - 3 \log[-\frac{1}{a}]^2 + 3 \log[1/x]^2)\label{poly2inf}
\eea
For the terms $-\pi^2 - 3 \log[-\frac{1}{a}]^2$ in (\ref{poly2inf}), it seems that we can set them to zero as this will lead to correct answer. We just throw them out by hand. Therefore we find a rule before taking the limit of $x\to\infty$,
\be
PolyLog[2,x/a]\to -\log[x] \log[1 - x/a]+\frac{1}{2}\log[x]^2\label{repl1}
\ee
\item $p=2$.
\be
G[2,1;x]=\log[x]^2 \log[1 - x/a] + 2 \log[x] PolyLog[2, x/a] - 2 PolyLog[3, x/a]
\ee
and the limit behaviour is
\bea
\lim_{x\to0}G[2,1;x]&=&0\\
\lim_{x\to\infty}G[2,1;x]&=&\frac{1}{3}(\pi^2 \log[-\frac{1}{a}] + \log[-\frac{1}{a}]^3 - \log[1/x]^3)
\eea
The same reasoning as $p=1$ case leads us to a replacement rule when $x\to\infty$,
\be
PolyLog[3,x/a]\to\frac{1}{3} \log[x]^3 - \frac{1}{2} \log[x]^2 \log[1 - x/a]\label{repl2}
\ee
\item $p=3$,
\bea
G[3,1;x]&=&\log[x]^3 \log[1 - x/a] + 3 \log[x]^2 PolyLog[2, x/a]\nn\\ &&-
 6 \log[x] PolyLog[3, x/a] + 6 PolyLog[4, x/a]
 \eea
 Then the replacement rule is
 \be
 PolyLog[4,x/a]\to\frac{1}{8}  \log[x]^4 - \frac{1}{6}\log[x]^3 \log[1 - x/a]\label{repl3}
 \ee

\end{enumerate}
The three replacement rules (\ref{repl1}),(\ref{repl2})and (\ref{repl3}) are all we need when we take the $x\to\infty$ limit.
\subsection*{Appendix B.1: Integral $F[J,j]$}
Here, we list the integral of $F[J,j]$ up to spin 6.
\begin{enumerate}
\item $J=3$. Since $j=0$ term is canceled, there are only two integrals.
\begin{enumerate}
\item
\be
F[3,1]=\frac{1}{12(U-1)^6}\sum_{i=0}^{2}c[3,1,i]\log^i[U]
\ee
with
\bea
c[3,1,0]&=&-(-1 + U)^2 (1 - 14 U + U^2)\nn\\
c[3,1,1]&=& 2(-1 + 8 U - 8 U^3 + U^4) \nn\\
c[3,1,2]&=& 12U^2\nn
\eea

\item
\be
F[3,2]=\frac{1}{2(U-1)^8}\sum_{i=0}^2c[3,2,i]\log^i[U]
\ee
with
\bea
c[3,2,0]&=&(-1 + U)^2 (5 + 26 U + 5 U^2)\nn\\
c[3,2,1]&=&-2 (-1 - 16 U + 16 U^3 + U^4)\nn\\
c[3,2,2]&=&4 U (2 + 5 U + 2 U^2)\nn
\eea
\end{enumerate}
\item $J=4,j=1,2,3$.
\begin{enumerate}
\item
\be
F[4,1]=\frac{1}{180 (-1 + U)^8}\sum_{i=0}^2c[4,1,i]\log^i[U]
\ee
with
\bea
c[4,1,0]&=&(-1 + U)^2 (2 - 23 U + 222 U^2 - 23 U^3 + 2 U^4) \nn\\
c[4,1,1]&=&-
 6 (-1 + 9 U - 45 U^2 + 45 U^4 - 9 U^5 + U^6) \nn\\
 c[4,1,2]&=&180 U^3\nn
\eea
\item
\be
F[4,2]=\frac{1}{36 (-1 + U)^{10}}\sum_{i=0}^2c[4,2,i]\log^i[U]
\ee
with
\bea
c[4,2,0]&=&-(-1 + U)^2 (11 - 248 U - 966 U^2 - 248 U^3 + 11 U^4)\nn\\
c[4,2,1]&=&
 6 (-1 + 18 U + 207 U^2 - 207 U^4 - 18 U^5 + U^6) \nn\\
c[4,2,2]&=&
 36 U^2 (9 + 22 U + 9 U^2)\nn
\eea
\item
\be
F[4,3]=\frac{1}{3 (-1 + U)^{12}}\sum_{i=0}^2c[4,3,i]\log^i[U]
\ee
with
\bea
c[4,3,0]&=&(-1 + U)^2 (10 + 209 U + 462 U^2 + 209 U^3 + 10 U^4)\nn\\
c[4,3,1]&=& -
 3 (-1 - 54 U - 189 U^2 + 189 U^4 + 54 U^5 + U^6)\nn\\
c[4,3,2]&=&9 U (3 + 24 U + 46 U^2 + 24 U^3 + 3 U^4)\nn
\eea
\end{enumerate}
\item $J=5;j=1,2,3,4$.
\begin{enumerate}
\item
\be
F[5,1]=\frac{1}{5040 (-1 + U)^{10}}\sum_{i=0}^2c[5,1,i]\log^i[U]
\ee
with
\bea
c[5,1,0]&=&-(-1 + U)^2 (9 - 110 U + 779 U^2 - 6396 U^3 + 779 U^4 - 110 U^5 +
    9 U^6)\nn\\
c[5,1,1]&=&
 12 (-3 + 32 U - 168 U^2 + 672 U^3 - 672 U^5 + 168 U^6 - 32 U^7 +
    3 U^8)\nn\\
c[5,1,2]&=&  5040 U^4\nn
\eea
\item
\be
F[5,2]=\frac{1}{360 (-1 + U)^{12}}\sum_{i=0}^2c[5,2,i]\log^i[U]
\ee
with
\bea
c[5,2,0]&=&(-1 + U)^2 (19 - 314 U + 4745 U^2 + 16300 U^3 + 4745 U^4 - 314 U^5 +
    19 U^6)\nn\\
c[5,2,1]&=& -
 12 (-1 + 16 U - 184 U^2 - 1776 U^3 + 1776 U^5 + 184 U^6 - 16 U^7 +
    U^8)\nn\\
c[5,2,2]&=&  720 U^3 (8 + 19 U + 8 U^2)\nn
\eea
\item
\be
F[5,3]=\frac{1}{72 (-1 + U)^{14}}\sum_{i=0}^2c[5,3,i]\log^i[U]
\ee
with
\bea
c[5,3,0]&=&-(-1 + U)^2 (31 - 1218 U - 17907 U^2 - 37412 U^3 - 17907 U^4 -
    1218 U^5 + 31 U^6) \nn\\
c[5,3,1]&=&
 12 (-1 + 32 U + 1232 U^2 + 3744 U^3 - 3744 U^5 - 1232 U^6 - 32 U^7 +
    U^8)\nn\\
c[5,3,2]&=&
 144 U^2 (18 + 128 U + 233 U^2 + 128 U^3 + 18 U^4) \nn
\eea
\item
\be
F[5,4]=\frac{1}{36 (-1 + U)^{16}}\sum_{i=0}^2c[5,4,i]\log^i[U]
\ee
with
\bea
c[5,4,0]&=&(-1 + U)^2 (141 + 6874 U + 42935 U^2 + 76500 U^3 + 42935 U^4 +
    6874 U^5 + 141 U^6) \nn\\
c[5,4,1]&=&-
 12 (-3 - 352 U - 3312 U^2 - 7008 U^3 + 7008 U^5 + 3312 U^6 +
    352 U^7 + 3 U^8)\nn\\
c[5,4,2]&=&
 144 U (4 + 66 U + 300 U^2 + 485 U^3 + 300 U^4 + 66 U^5 + 4 U^6) \nn
\eea
\end{enumerate}
\item $J=6;j=1,2,3,4,5$.
\begin{enumerate}
\item
\be
F[6,1]=\frac{1}{25200 (-1 + U)^{12}}\sum_{i=0}^2c[6,1,i]\log^i[U]
\ee
with
\bea
c[6,1,0]&=&(-1 + U)^2 (8 - 109 U + 774 U^2 - 4343 U^3 + 32540 U^4 - 4343 U^5 +
    774 U^6\nn\\&& - 109 U^7 + 8 U^8)\nn\\
c[6,1,1]&=& -
 20 (-2 + 25 U - 150 U^2 + 600 U^3 - 2100 U^4 + 2100 U^6\nn\\&& - 600 U^7 +
    150 U^8 - 25 U^9 + 2 U^{10})\nn\\
c[6,1,2]&=&25200 U^5 \nn
\eea
\item
\be
F[6,2]=\frac{1}{25200 (-1 + U)^{14}}\sum_{i=0}^2c[6,2,i]\log^i[U]
\ee
with
\bea
c[6,2,0]&=&-(-1 + U)^2 (261 - 4228 U + 42408 U^2 - 538956 U^3 - 1720570 U^4 -
    538956 U^5 \nn\\&&+ 42408 U^6 - 4228 U^7 + 261 U^8) \nn\\
c[6,2,1]&=&
 60 (-3 + 50 U - 475 U^2 + 4400 U^3 + 37800 U^4 - 37800 U^6 -
    4400 U^7\nn\\&& + 475 U^8 - 50 U^9 + 3 U^{10})\nn\\
c[6,2,2]&=& 25200 U^4 (25 + 58 U + 25 U^2)\nn
\eea
\item
\be
F[6,3]=\frac{1}{1800 (-1 + U)^{16}}\sum_{i=0}^2c[6,3,i]\log^i[U]
\ee
with
\bea
c[6,3,0]&=&(-1 + U)^2 (137 - 3601 U + 94286 U^2 + 1146173 U^3 + 2288810 U^4 +
     1146173 U^5\nn\\&& + 94286 U^6 - 3601 U^7 + 137 U^8) \nn\\
c[6,3,1]&=&-
  60 (-1 + 25 U - 525 U^2 - 16400 U^3 - 45100 U^4 + 45100 U^6 \nn\\&&+
     16400 U^7 + 525 U^8 - 25 U^9 + U^{10}) \nn\\
c[6,3,2]&=&3600 U^3 (50 + 325 U + 573 U^2 + 325 U^3 + 50 U^4)\nn
\eea
\item
\be
F[6,4]=\frac{1}{360 (-1 + U)^{18}}\sum_{i=0}^2c[6,4,i]\log^i[U]
\ee
with
\bea
c[6,4,0]&=&-(-1 + U)^2 (187 - 11376 U - 382064 U^2 - 2076752 U^3 - 3527190 U^4 -
    2076752 U^5\nn\\&& - 382064 U^6 - 11376 U^7 + 187 U^8)\nn\\c[6,4,1]&=&
 60 (-1 + 50 U + 4175 U^2 + 32800 U^3 + 62800 U^4 - 62800 U^6\nn\\&& -
    32800 U^7 - 4175 U^8 - 50 U^9 + U^{10})\nn\\ c[6,4,2]&=&
 3600 U^2 (10 + 140 U + 575 U^2 + 902 U^3 + 575 U^4 + 140 U^5 +
    10 U^6)\nn
\eea
\item
\be
F[6,5]=\frac{1}{180 (-1 + U)^{20}}\sum_{i=0}^2c[6,5,i]\log^i[U]
\ee
with
\bea
c[6,5,0]&=&(-1 + U)^2 (786 + 70697 U + 873858 U^2 + 3485019 U^3 + 5427680 U^4 +
    3485019 U^5 \nn\\&&+ 873858 U^6 + 70697 U^7 + 786 U^8) \nn\\c[6,5,1]&=&-
 60 (-3 - 625 U - 11125 U^2 - 56500 U^3 - 89250 U^4 + 89250 U^6 \nn\\&&+
    56500 U^7 + 11125 U^8 + 625 U^9 + 3 U^{10})\nn\\c[6,5,2]&=&
 900 U (5 + 140 U + 1160 U^2 + 3820 U^3 + 5626 U^4 + 3820 U^5 +
    1160 U^6 + \nn\\&&140 U^7 + 5 U^8)\nn
\eea
\end{enumerate}
\end{enumerate}
\subsection*{Appendix B.2: Integral in $\mathcal{O}(\mu^4)$}
The basic integral is
\be
(\prod_{k=1}^4\int dt_k f[J,t_k])y_{12}^ay_{i3}^by_{j4}^c.
\ee
In this work, $a,b,c$ are positive integers and $1\le i\le2,1\le j\le3$. It is convenient to define a set of integrals as\footnote{We omit the dependence of J to simplify notation. In the following, our calculation is done for $J=3$.}
\bea
m_a[t]&=&\int_{0}^{\infty}dt' f[t']y[t,t']^a\nn\\
g_{ab}[t]&=&\int_{0}^{\infty}dt'f[t']m_b[t']y[t,t']^a\nn\\
k_{abc}[t]&=&\int_{0}^{\infty}dt'f[t']m_b[t']m_c[t']y[t,t']^a\nn\\
l_{abc}[g]&=&\int_{0}^{\infty}dt'f[t']g_{bc}[t']y[t,t']^a.\nn
\eea
Then after integrating out $t_4,t_3,t_2$, the terms $y_{12}^ay_{i3}^by_{j4}^c$ are replaced by
\bea
y_{12}^ay_{13}^by_{14}^c&\to& m_a[t_1]m_b[t_1]m_c[t_1]\nn\\
y_{12}^ay_{23}^by_{14}^c&\to& g_{ab}[t_1]m_c[t_1]\nn\\
y_{12}^ay_{13}^by_{24}^c&\to& g_{ac}[t_1]m_b[t_1]\nn\\
y_{12}^ay_{23}^by_{24}^c&\to& k_{abc}[t_1]\nn\\
y_{12}^ay_{13}^by_{34}^c&\to& m_a[t_1]g_{bc}[t_1]\nn\\
y_{12}^ay_{23}^by_{34}^c&\to& l_{abc}[t_1].\nn
\eea
So the building blocks are these defined functions. One can integrate out them term by term, though very tedious. In our computation, we use a slightly different method. Instead of integrating out $t_4,t_3,t_2$ term by term, we just integrate out $t_4,t_3$ at first. And then we sum over the results. After that, the integral $t_2, t_1$ are done.  That means we just do the following replacement,
\bea
y_{13}^by_{14}^c&\to& m_{b}[t_1]m_{c}[t_1]\nn\\
y_{23}^by_{14}^c&\to& m_b[t_2]m_c[t_1]\nn\\
y_{13}^by_{24}^c&\to& m_b[t_1]m_c[t_2]\nn\\
y_{23}^by_{24}^c&\to& m_b[t_2]m_c[t_2]\nn\\
y_{13}^by_{34}^c&\to& g_{bc}[t_2]\nn\\
y_{23}^by_{34}^c&\to&g_{bc}[t_2].\nn
\eea
The relevant $m_a$s are the ones with $a=1,2,\cdots,5$, and the relevant $g_{bc}$ are
\be
 b+c\le6,\  b\ge1,\ c\ge1
 \ee
 So the number of integrals we need to do is $(5+5+4+3+2+1)=$20.

Then we classify the function according to the power of $n$ and $y_{12}$. A general function is
\be
n^{2\alpha+1}y_{12}^a\mathcal{F}_{\alpha,a}(t_1,t_2)/n^8\label{cla}
\ee
where $\mathcal{F}_{\alpha,a}(t_1,t_2)$ is determined by $m_a,g_{ab}$. $\alpha,a$ satisfy
\be
1\le\alpha\le4,\ 1\le a\le 2\alpha+2
\ee
So there are $(4+6+8+10=)28$ different integrals of $t_2$ in (\ref{cla}).
In the work, we calculate the $\mathcal{O}(c)$ part and the quantum correction, there is no divergence after we sum over the integrals of $t_4,t_3,t_2$.

After all these have been done, we need to integrate $t_1$. We can integrate it according to the power of n. So there are 4 integrals of $t_1$. Since the R$\acute{e}$nyi entropy at this order should be finite, the integration should cancel the divergent term in the partition function $n\log Z_1$. So the divergence should come from $\alpha=4$ terms. For $\alpha=4$, we should be careful to seperate the divergence term. There is no other subtlety except these. In total, we need to do $(28+20+4=)$52 integrals. We just mention that the first $28$ and the last 4 integrals is relatively simple, whilst the other 20 integrals of $t_2$ is a bit complicated. As an illustration, we just give the results of $m_1[t]$, the other functions $m_a[t],g_{ab}[t]$ are similar.
\bea
m_1[t]&=&\frac{1}{-1 + U}(-1 + t) (-\frac{1}{2 (-1 + t) (-1 + U)^2} -
  \frac{ U}{(-1 + U)^3 (-t + U)} - \frac{
   1 + U - 2 t U}{(-1 + t)^2 (-1 + U)^3} \nn\\&&-\frac{
   t^2 \log[t]}{(-1 + t)^3 (t - U)^2} + \frac{
   U (-t (2 + U) + U (1 + 2 U)) \log[U]}{(t- U)^2 (-1 + U)^4}).
\eea

\section*{Appendix C: $\mathcal{O}(\mu^2)$ Correction to R$\acute{e}$nyi Entropy and Entanglement Entropy of Other Spins}
The formula for R$\acute{e}$nyi entropy to $\mathcal{O}(\mu_J^2)$ is (\ref{regs}). One can plug the value of $\tilde{b}[J,j;n]$ and $F[J,j]$ to find their explicit expression. Taking the limit $n\to1$, we can list the $\mathcal{O}(\mu_J^2)$ corrections of entanglement entropies
from spin 4 to spin 6 below,
\bea
S|_{spin 4,\mu_4^2}&=&\frac{8\mu_4^2\mathcal{N}_4\pi^6}{105\beta^4(U-1)^6}\sum_{i=0}^2 \kappa[4,i]Log^i[U],\\
S|_{spin 5,\mu_5^2}&=&-\frac{32\mu_5^2\mathcal{N}_5\pi^8}{2835\beta^6(U-1)^8}\sum_{i=0}^2\kappa[5,i]Log^i[U],\\
S|_{spin 6,\mu_6^2}&=&\frac{64\mu_6^2\mathcal{N}_6\pi^{10}}{10395\beta^8(U-1)^{10}}\sum_{i=0}^2\kappa[6,i]Log^i[U]
\eea
with
\bea
\kappa[4,0]&=&-5 (5 + 27 U^2 - 64 U^3 + 27 U^4 + 5 U^6)\\
\kappa[4,1]&=&
  6 (-3 - 8 U - 35 U^2 + 35 U^4 + 8 U^5 + 3 U^6) \\
  \kappa[4,2]&=&-
  36 U (1 + U + 6 U^2 + U^3 + U^4)
\eea
\bea
\kappa[5,0]&=&-(-1 + U)^2 (213 + 982 U + 7175 U^2 + 8460 U^3 + 7175 U^4 + 982 U^5 +
     213 U^6)
     \\
     \kappa[5,1]&=&
  24 (-6 - 29 U - 279 U^2 - 381 U^3 + 381 U^5 + 279 U^6 + 29 U^7 +
     6 U^8)\\
     \kappa[5,2]&=&  -
  360 U (1 + 3 U + 21 U^2 + 20 U^3 + 21 U^4 + 3 U^5 + U^6)
\eea
\bea
\kappa[6,0]&=&-(-1 + U)^2 (463 + 3751 U + 43414 U^2 + 119077 U^3 + 195790 U^4 +
     119077 U^5\nn\\&& + 43414 U^6 + 3751 U^7 + 463 U^8) \\
     \kappa[6,1]&=&
  60 (-5 - 37 U - 597 U^2 - 1972 U^3 - 2912 U^4 + 2912 U^6 +
     1972 U^7 + 597 U^8\nn\\&& + 37 U^9 + 5 U^{10})\\
     \kappa[6,2]&=& -
  900 U (1 + 6 U + 56 U^2 + 126 U^3 + 210 U^4 + 126 U^5 + 56 U^6 +
     6 U^7 + U^8)
\eea

\section*{Appendix D: Quantum Correction of Partition Function of Higher Spin Black Hole}
In this Appendix, we use the zero mode insertion method \cite{Gaberdial1203} to calculate the quantum correction of the partition function of spin 3 black hole. We will find the same result as section 5.  In their method, the partition function is supposed to be\footnote{In \cite{Gaberdial1203}, $\hat{\tau}$ is the parameter $\tau$ in this paper. Only in this Appendix, we change our convention to match those in \cite{Gaberdial1203}. The parameter $\tau$ in this Appendix is $\tau=-\frac{1}{\tilde{\tau}}$. }
 \be
Z_{CFT}(\hat{\tau},\alpha)=Tr \hat{q}^{L_0-\frac{c}{24}}y^{W_0}, \hat{q}=e^{2\pi i\hat{\tau}},y=e^{2\pi i\alpha}
\ee
In the perturbation theory, the partition function is expanded by the power of $\alpha$,
\be
Z_{CFT}(\hat{\tau},\alpha)=Tr(\hat{q}^{L_0-\frac{c}{24}})+\frac{(2\pi i\alpha)^2}{2}Tr(W_0^2\hat{q}^{L_0-\frac{c}{24}})+\frac{(2\pi i\alpha)^4}{4!}Tr(W_0^4\hat{q}^{L_0-\frac{c}{24}})+\cdots
\ee
In the high temperature regime, after a  S modular transformation, the trace  is contributed by vacuum state.  There is no quantum correction at order $\alpha^2$, hence we proceed to the $\alpha^4$ correction. Borrowing the notation in \cite{Gaberdial1203},
\be
Z^{(4)}=\frac{\alpha^4\tau^8}{4!}(3I_1(A_1+A_2)+3I_2A_6+\frac{5}{2}I_3A_3+\frac{5}{3}I_4A_4+I_5A_5)
\ee
As they ignore the subleading order in 1/c, we denote their result (3.19) and (3.20) by $A_i^{cl}$. Schematically, we can write
\be
A_i=A_i^{cl}+A_i^{qu}
\ee
To read out the quantum correction, we consider the effect from the composite operator $\Lambda=:TT:(z)-\frac{3}{10}\partial^2T(z)$. In their convention, it is easy to find
\be
[\Lambda_m,\Lambda_n]=\frac{c(5c+22)}{10\times 7!}m(m^2-1)(m^2-4)(m^2-9)\delta_{m,-n}+\cdots
\ee
The $\cdots$ term has no contribution at order $\alpha^4$. We define some notation,
\bea
W[0]&=&(2\pi i)^{-1}\sum_{j\ge-2}a_jW_j\nn\\
W[1]&=&(2\pi i)^{-2}\sum_{j\ge-1}b_jW_j\nn\\
W[2]&=&(2\pi i)^{-3}\sum_{j\ge0}c_jW_j\nn\\
W[3]&=&(2\pi i)^{-4}\sum_{j\ge1}d_jW_j
\eea
The values of $a_j,b_j,c_j,d_j$ can be read from \cite{Gaberdial1203}, Appendix B.1.  We use $A_1$ as an example.
\bea
A_1&=&b_5b_{-1}b_{-1}<W_5W_{-1}W_{-1}W_{-3}>+b_4b_0b_{-1}<W_4W_0W_{-1}W_{-3}>+b_4b_{-1}b_0<W_4W_{-1}W_0W_{-3}>+\nn\\
&&b_3b_0b_0<W_3W_0W_0W_{-3}>+b_3b_{-1}b_1<W_3W_{-1}W_1W_3>+b_3b_1b_{-1}<W_3W_1W_{-1}W_{-3}>
\eea
We note that in each term above, it is actually the vacuum expectation value of two commutator, for example
\be
<W_5W_{-1}W_{-1}W_{-3}>=<[W_5,W_{-1}][W_{-1},W_{-3}]>
\ee
Using the nonlinear $W_{\infty}(\lambda)$ algebra
\be
[W_m,W_n]\sim U_{m+n}+L_{m+n}+\delta_{m,-n}+\frac{40N_3}{5c+22}(m-n)\Lambda_{m+n},
\ee
The first three terms in the right hand side contribute $A_{i}^{cl}$, whist the last term contributes $A_i^{qu}$.
We find
\be
A_1^{qu}=-\frac{416 c N_3^2}{63 (22 + 5 c)}=A_5^{qu},
A_2^{qu}=-\frac{32 c N_3^2}{7 (22 + 5 c)}=A_3^{qu},
A_4^{qu}=-A_6^{qu}=3/2A_2^{qu}.
\ee
Substituting the values of $I_i,N_3,a_j,b_j,c_j,d_j$ in that paper, we find the quantum correction of partition function at $\alpha^4$,
\be
\log Z|_{\mu^4,quan}=\frac{640 i \alpha^4 c \pi}{27 (22 + 5 c)\hat{\tau}^9},\label{q}
\ee
Note (\ref{q}) is exactly the same as (\ref{qpf}).

\end{document}